\pdfoutput=1
\documentclass[letterpaper,twocolumn,10pt]{article}
\usepackage[twoside=true, head=13pt,
     paperwidth=8.5in, paperheight=11in,
     includeheadfoot=false, columnsep=2pc,
     top=1in, bottom=1in, inner=0.75in, outer=0.75in,
     marginparwidth=2pc,heightrounded]{geometry}

\usepackage{graphicx}
\usepackage[tt=false, type1=true]{libertine}
\usepackage{mathptmx}
\usepackage{amssymb}
\usepackage[varqu]{zi4}
\usepackage[T1]{fontenc}
\usepackage{enumitem}
\usepackage[font={footnotesize},labelfont={footnotesize,bf},textfont={footnotesize,it}]{caption}
\usepackage{xspace}
\usepackage{xcolor}
\usepackage{booktabs}
\usepackage{tabularx}
\usepackage{array}
\usepackage{subcaption}
\usepackage{amsmath}
\usepackage{url}
\usepackage{longtable}
\usepackage{listings}
\usepackage{pifont}
\usepackage{authblk}

\newcommand{\cmark}{\ding{51}}
\newcommand{\xmark}{\ding{55}}

\newcommand{\specexcerpt}[1]{%
  \fcolorbox{gray!55}{gray!8}{%
    \begin{minipage}{\dimexpr\columnwidth-2\fboxsep-2\fboxrule\relax}
    \scriptsize\setlength{\parindent}{0pt}#1
    \end{minipage}%
  }%
}

\lstdefinestyle{prompt}{
  basicstyle=\ttfamily\scriptsize,
  breaklines=true,
  breakindent=0pt,
  columns=fullflexible,
  keepspaces=true,
  frame=single,
  rulecolor=\color{gray!50},
  backgroundcolor=\color{gray!7},
  framesep=4pt,
  xleftmargin=4pt,
  xrightmargin=2pt,
  aboveskip=4pt,
  belowskip=4pt,
  showstringspaces=false,
}

\newcommand{\sys}{\textsc{SpecLens}\xspace}

\newcommand{\paragraphb}[1]{\noindent\textbf{#1}\quad}

\newenvironment{denseitemize}{
\begin{itemize}[topsep=2.5pt, partopsep=0pt, leftmargin=1.5em]
  \setlength{\itemsep}{2.5pt}
  \setlength{\parskip}{0pt}
  \setlength{\parsep}{0pt}
}{\end{itemize}}

\newif\ifdraft\draftfalse
\ifdraft
\newcommand{\zt}[1]{\textcolor{teal}{Zhaowei: {#1}}}
\newcommand{\ziyue}[1]{\textcolor{purple}{Ziyue: {#1}}}
\newcommand{\sxt}[1]{\textcolor{cyan}{Sixu: {#1}}}

\else
\newcommand{\zt}[1]{}
\newcommand{\ziyue}[1]{}
\newcommand{\sxt}[1]{}

\fi

\title{Can LLMs help find Ambiguities in Protocol Specifications?}

\author[1]{Ziyue Dang\thanks{Equal contribution (co-first authors).}}
\author[2]{Sixu Tan\protect\footnotemark[1]}
\author[2]{Atharva Nevasekar}
\author[2]{Zhaowei Tan}
\author[1]{George Varghese}
\author[1]{Songwu Lu}
\affil[1]{UCLA\protect\\ \texttt{\{ziyue.dang, varghese, slu\}@cs.ucla.edu}}
\affil[2]{University of California, Riverside\protect\\ \texttt{\{stan073, aneva018, ztan\}@ucr.edu}}
\date{}

\begin{document}
\pagestyle{plain}
\maketitle
\begin{abstract}
Internet protocol specifications written in RFCs are subject to ambiguities and multiple interpretations that can cause interoperability failure. While these have presumably cleared up after years of experience, such ambiguities can
bedevil the adoption of newer protocols like 5G. The 5G specifications pair a formal message syntax (ASN.1) with message-handling procedures written in natural language. This creates semantic underspecification: a syntactically valid message can reach a state whose procedures never say how to handle it, so standard-compliant implementations diverge. We frame this as a gap or a fork in a partially specified communicating state machine, and present \sys, which puts that view in front of a language model as a scaffold. Stronger models do not remove the need for it: they broaden the search without disciplining it, and fewer than half their findings survive inspection. Across 36 procedures from six 3GPP and O-RAN protocols, experts accept 185 of 197 \sys findings, and 60 drive observable divergence between the OpenAirInterface and srsRAN implementations under differential test. While we use 5G as a canonical example of a newer protocol, we also show ambiguity results for the more mature DNS protocol.
\end{abstract}

\section{Introduction}
\label{sec:intro}

Early protocol standards, exemplified by ISO and ITU, relied on complete specifications (ideally in formal languages) designed before implementation. By contrast, the Internet relied on ``rough consensus and running code''~\cite{clark} and English language specifications in RFCs.  While these were wildly successful, the historic axiom ``be conservative in what you send, be liberal in what you accept'' often masked specification flaws. These sometimes surfaced as attack vectors that required implementation changes: for example, overlapping IP fragments~\cite{teardrop} and malformed IP options~\cite{OSfingerprinting}.  
In some instances, such as the TOS field, which was originally loosely designed to let applications specify network preferences, the IETF completely rewrote the rules multiple times~\cite{rfc1349,rfc2474,rfc3168}. These mid-lifecycle updates caused discrepancies between legacy network hardware and modern equipment. This raises the question: {\em is there a better way to find specification ambiguities early for newer protocols} without requiring protocol designers to abandon English prose specifications instead of hard to learn formal languages such as LOTOS~\cite{lotos}, Estelle~\cite{estelle}, or SDL~\cite{sdl}.

We ground our study of protocol ambiguity in the more recent 5G protocols that are arguably less mature than the classic Internet suite.  The 5G system provides mobile internet access for billions of users. Its design has been standardized by the 3GPP (3rd Generation Partnership Project)~\cite{3gpp} and O-RAN (Open Radio Access Network)~\cite{oran_alliance} specifications. These specifications define both the message formats (syntax) and the message handling procedures written in natural language. At the end of the paper we also turn our lens briefly on a very mature protocol, DNS, and find (perhaps surprisingly) several unresolved ambiguities.

\begin{figure*}[t]
  \centering
  \begin{subfigure}{0.45\textwidth}
    \centering
    \includegraphics[width=\linewidth]{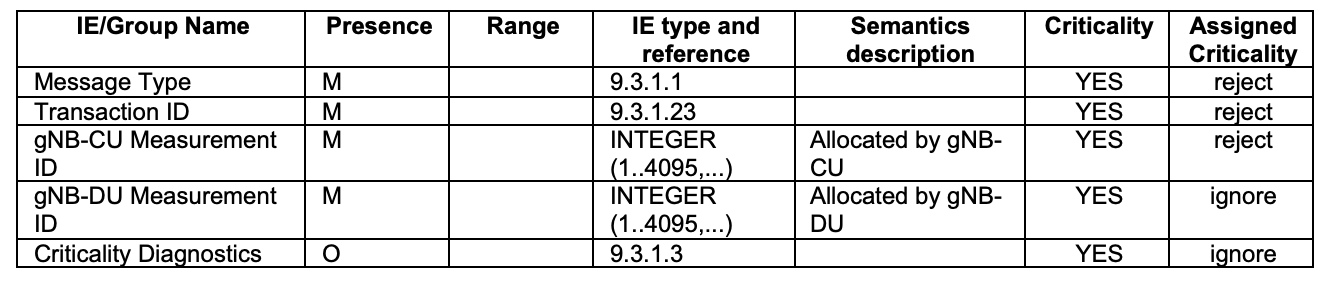}
    \caption{An information-element (IE) syntax table.}
    \label{fig:syntax-table}
  \end{subfigure}%
  \hfill
  \begin{subfigure}{0.53\textwidth}
    \centering
    \includegraphics[width=\linewidth]{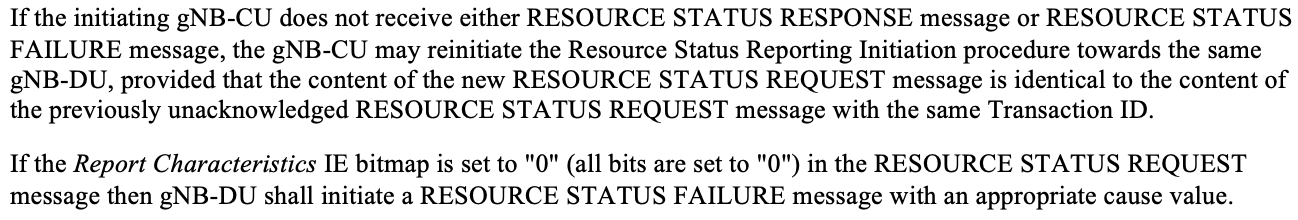}
    \caption{A natural-language procedure description.}
    \label{fig:proc-text}
  \end{subfigure}
  \vspace{-0.05in}
  \caption{The two parts of a 3GPP/O-RAN specification.}
  \label{fig:3gpp-examples}
\end{figure*}

More precisely, we study \emph{semantic underspecification}: procedural text that fails to guide the processing of syntactically valid messages. Note that our problem is not about identifying specification errors (where erroneous descriptions are given) or implementation bugs (where code fails to meet requirements), both studied extensively in prior work~\cite{klischies2023instructions,cellularlint,pawagi2025rfcscope,contester}.

Since protocols are defined in natural language, Large Language Models (LLMs) are conceptually promising, given their strengths in semantic understanding and reasoning. However, we find that a naïve direct prompting approach does not work. It rediscovers a few shallow, high-probability gaps across many fields and misses many cross-step cases. 

{\em Will more powerful models solve the problem?} Many software engineering scaffolds became redundant with the advent of Claude 4.5.  
We found experimentally (\S\ref{sec:eval:rq1}) that removing the scaffold altogether by giving whole specifications to frontier agentic models (e.g., Opus 5.0), buys breadth but not discipline. Fewer than half the findings mined without our scaffold survive inspection, and a third conflate freedom the standard purposely grants with specification failure. Further, a third of our scaffolded findings are missed without a scaffold. Worse, without scaffolding, the model output is an unstructured grab bag of underspecification candidates, riddled with false positives and negatives.

Instead, we propose \sys, a scaffolded LLM-driven analysis framework that systematically uncovers underspecifications in 5G protocols. \sys works with available LLM tools, but departs from conventional end-to-end prompting. {\em Our key contribution is to model a specification as a set of communicating state machines and to use this model to infer a set of underspecification categories that flow directly from the state machine definition.}  

We then orchestrate a staged inference workflow using our state machine lens: context construction, taxonomy-based mining, gap/fork triage, and impact analysis with a deterministic state-machine engine that also derives a differential test (\S\ref{sec:design}). Rather than replacing domain experts, \sys produces a ranked set of under-specified cases with textual evidence, an impact estimate, and a test design.

Our preliminary evaluation covers 36 procedures across six 3GPP and O-RAN protocols, run with three LLM backends. The findings hold up under scrutiny: reading each candidate against the standard, two domain experts conservatively support 185 of 197 proposals, and 195 are supported by code or at least one expert. 60 of the findings drive observable divergence between the independent OpenAirInterface and srsRAN stacks under differential test, most of them incompatible decisions such as one implementation accepting where the other rejects. Some are outright hazards: a single legal \textsf{PDU SESSION RESOURCE RELEASE COMMAND} carrying one stale session identifier frees a single session on one implementation but tears down the device's entire connection on the other.

We also describe early results with the more mature (50+ years) DNS, finding, perhaps surprisingly, a few: e.g., how much to advance a zone's SOA serial number on an update (RFC~2136~\cite{rfc2136}).

Our results in 5G and DNS open up three directions (\S\ref{sec:discussion}). Several validated underspecifications are already exploitable: this motivates mining using an adversarial lens rather than an interoperability one. Second, our unit of analysis is one document, while a modern protocol's contract is spread across many, so the ambiguities that live in the interplay between documents are ones we cannot yet express. Finally, communicating state machines are one model among several: a richer model such as timed automata~\cite{alur1994timed} may produce a finer-grain taxonomy at the cost of harder extraction.

\section{Background}
\label{sec:background}

The 5G system is defined by two standards bodies. The 3rd Generation Partnership Project (3GPP) specifies the radio access network (RAN) and its core-network protocols~\cite{3gpp}, and the O-RAN Alliance specifies an open, disaggregated realization of that RAN~\cite{oran_alliance}. Disaggregation divides the RAN into components that may come from different vendors, such as a central unit (CU) and one or more distributed units (DUs), and connects them over standardized interfaces: F1 between the CU and a DU, E1 inside a split CU, NG between the RAN and the core, and E2 between the RAN and its near-real-time controller. These interfaces are served by dedicated protocols: F1AP~\cite{3gpp-ts-38.473}, E1AP~\cite{3gpp-ts-38.463}, NGAP~\cite{3gpp-ts-38.413}, and E2AP~\cite{ORAN-E2AP} for control-plane signaling, GTP-U~\cite{3gpp-ts-29.281} for the user plane, and E2 service models such as E2SM-KPM~\cite{ORAN-E2SM-KPM} for the content exchanged over E2. Because independently built components must interoperate across these interfaces, the interface specifications carry the full weight of defining compatible behavior.

A specification fixes that behavior in two parts, shown in Figure~\ref{fig:3gpp-examples}. The \textbf{syntax}, expressed in ASN.1, defines the format of each message and the information elements (IEs) it may carry (Figure~\ref{fig:syntax-table}). The \textbf{procedures}, written in natural language, describe how a node processes a received message and how it updates its internal state (Figure~\ref{fig:proc-text}). Syntax and procedures together fix a message's \textbf{semantics}: the action a node takes and the state it reaches upon receiving that message. These semantics are state-dependent. The same syntactically valid message can demand different handling according to the receiver's \textbf{protocol state}, which comprises its active connections, pending operations, and the history of messages exchanged so far.

\section{The Problem of Underspecification}
\label{sec:problem}

\begin{figure}[t]
  \centering
  \includegraphics[width=\columnwidth]{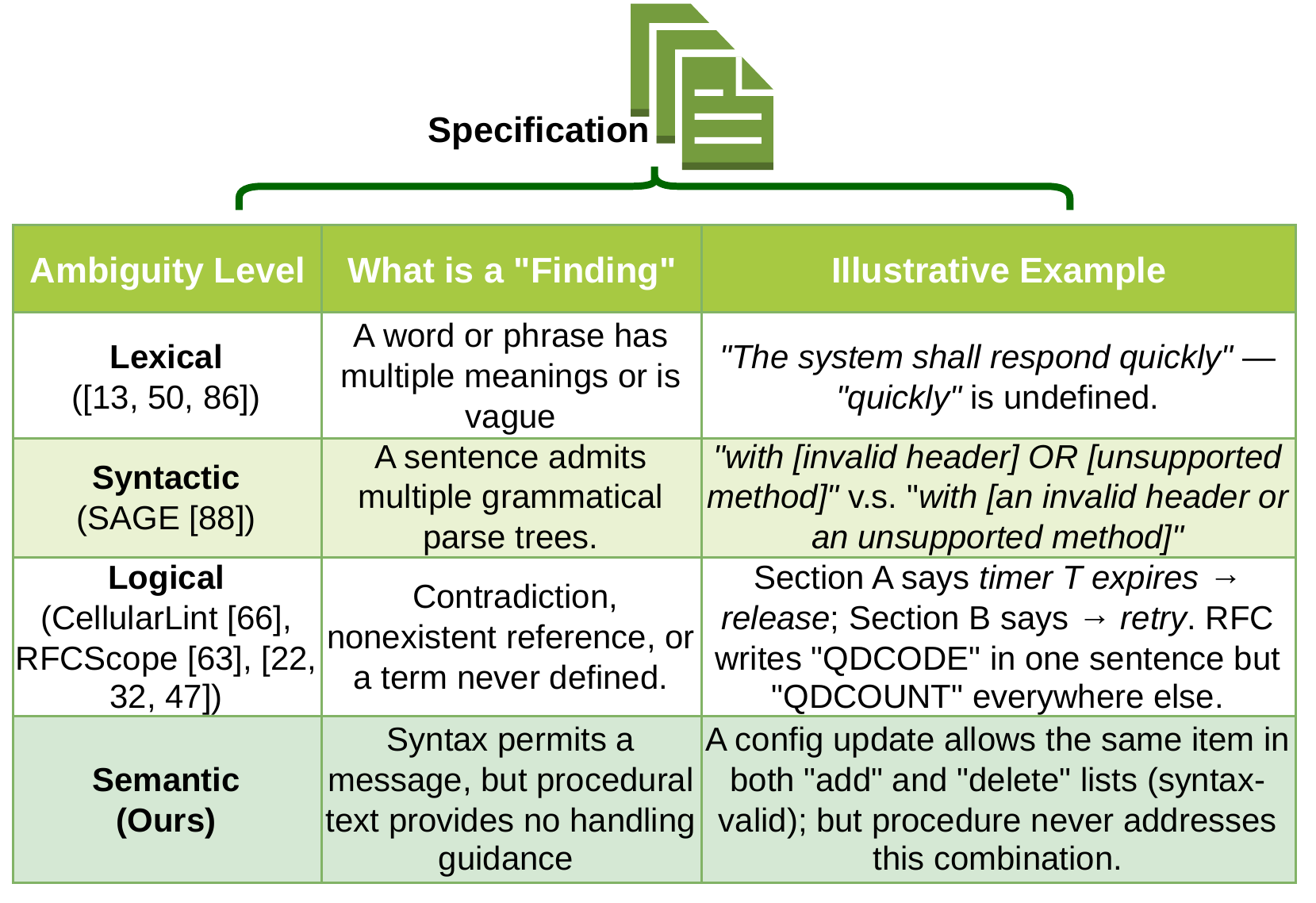}
  \vspace{-0.15in}
  \caption{Where semantic underspecification sits among prior ambiguities.}
  \label{fig:underspec_comparison}
\end{figure}

We study \textbf{semantic underspecification}: cases in which a specification admits a syntactically valid message in a reachable protocol state, yet its natural-language procedures do not determine how the receiver should handle it. Such a case is not an error. Each individual statement may be correct; the specification is simply silent, or admits several incompatible readings, about a scenario its own syntax permits. This sets our target apart from most prior work on specification quality, which looks for a defect that is present in the text or in a model built from it: a contradiction or broken cross-reference between clauses~\cite{cellularlint,pawagi2025rfcscope}, a word or phrase open to more than one meaning~\cite{wildenburg2024pre,mahbub2024can}, or a transition missing from a state machine an analyst has already built by hand~\cite{klischies2023instructions}. Figure~\ref{fig:underspec_comparison} places these notions on a common scale. Semantic underspecification sits at the boundary between a specification's formal syntax and its prose, where the two are together meant to fix behavior but do not.

Underspecification is a structural consequence of how these documents are written, not an accident of any one of them. Message syntax is defined formally in ASN.1, admitting an enormous space of valid messages and field combinations. The procedures that give those messages meaning are prose, which cannot enumerate every combination of message and state. What remains is a residue of syntactically valid scenarios for which the procedures prescribe no clear behavior. In a single-vendor deployment such a scenario may never surface; across the multi-vendor open interfaces of \S\ref{sec:background} it becomes an interoperability hazard~\cite{samsung_multivendor_open_ran, ntia_5g_challenge_tm23568}, because two independently built components can resolve the same silent case in incompatible ways. The residue also grows rather than shrinks, since each release adds messages, fields, and interfaces faster than prose can cover their combinations.

\subsection{A state-machine view of specifications}
\label{sec:problem:cfsm}

To make ``does not determine how the receiver should handle it'' precise, we model a protocol as a set of communicating state machines, one per role, that exchange messages over channels~\cite{brand1983cfsm}. Each machine has a set of local states and a set of transitions. A transition fires when the machine receives a message $m$ in a state $s$; it moves the machine to a next state $s'$ and may emit messages of its own. We write it as $(s, m) \rightarrow (s', o)$, with the current state and input on the left and the resulting state and output on the right.

A complete protocol would fix exactly one such transition for every reachable state and every message that can arrive there. A natural-language specification does not. It induces a transition relation $\delta^{\mathrm{spec}}$ that may leave a case out or permit more than one resolution, so the behaviors it allows form a \emph{set} of runs $R^{\mathrm{spec}}$ rather than a single one. A conformant implementation is any deterministic choice inside that set, a refinement $\delta^{\mathrm{impl}} \subseteq \delta^{\mathrm{spec}}$, so two conformant implementations can differ wherever $\delta^{\mathrm{spec}}$ fails to pin the transition down. Underspecification is that freedom.

\begin{figure}[t]
  \centering
  \specexcerpt{%
    \textbf{8.3.2.1\quad General.} The purpose of the UE Context Release Request procedure is to enable the gNB-DU to request the gNB-CU to release the UE-associated logical F1-connection [\ldots].

    \smallskip
    \textbf{8.3.2.2\quad Successful Operation.} The gNB-DU [\ldots] initiates the procedure by generating a UE CONTEXT RELEASE REQUEST message towards the affected gNB-CU [\ldots]. The UE Context Release procedure \emph{may} be initiated upon reception of a UE CONTEXT RELEASE REQUEST message.

    \smallskip
    \textbf{8.3.2.3\quad Abnormal Conditions.} If one or more candidate cells in the Candidate Cells To Be Cancelled List IE [\ldots] were not prepared using the same UE-associated signaling connection, the gNB-CU shall ignore those non-associated candidate cells.%
  }
  \vspace{-0.04in}
  \caption{F1AP UE Context Release Request (TS~38.473~\S8.3.2): the Abnormal Conditions clause covers only stray candidate cells, not a request naming a connection the gNB-CU does not hold.}
  \label{fig:f1ap-excerpt}
\end{figure}

The F1AP UE Context Release Request procedure shows how that freedom arises (Figure~\ref{fig:f1ap-excerpt})~\cite{3gpp-ts-38.473}. A gNB-DU asks the gNB-CU to release a UE's F1 connection by sending a \textsf{UE CONTEXT RELEASE REQUEST}, upon which the gNB-CU may initiate the release. The procedure constrains just one edge case, stray candidate cells, and is silent on another: a request that names a UE-associated connection the gNB-CU does not currently hold. That configuration is reachable and the message can arrive in it, yet no transition is defined there. Release handling underspecified in this way is not hypothetical: \S\ref{sec:eval:rq3} shows that gaps around releasing a context that is not currently active drive observable divergence between independent gNB implementations.

We can state the condition exactly. Fix a reachable state $s$ and a message $m$ that can arrive there, and let $\Delta^{\mathrm{spec}}(s,m)$ be the set of next states the specification allows. The scenario is well specified when this set holds exactly one element. It is underspecified in one of two forms: $|\Delta^{\mathrm{spec}}(s,m)| = 0$, a \textbf{gap}, where no transition is given, or $|\Delta^{\mathrm{spec}}(s,m)| \ge 2$, a \textbf{fork}, where several are. The case above is a gap. We return to what the two forms imply in \S\ref{sec:problem:gapfork}, and give the full run semantics, together with a test that separates underspecification from questions the model cannot answer, in Appendix~\ref{app:model}.

\subsection{Where underspecification lives}
\label{sec:problem:categories}

The case above under-constrains a single transition, but not every opening is local. A protocol's behavior is a set of runs, and a run is an arrangement of transition firings, each with its own content. Underspecification can therefore enter at either level: in how firings are arranged across a run, or in what one firing consumes and produces. These levels split into four components, which structure the search in \S\ref{sec:design:mining}.

\paragraphb{Local content.} A transition's \textbf{trigger} is the current state together with the message that enables it; its \textbf{effect} is the next state and any output it produces. A \emph{trigger} is underspecified when it is unclear whether a transition applies at all in a given state on a given message. An \emph{effect} is underspecified when the message is accepted but its consequences, the next state, the response, or the cleanup, are left open.

\paragraphb{Temporal arrangement.} Across several firings, two questions arise that no single transition answers; they are the safety and liveness sides into which a run property classically decomposes~\cite{alpern1985defining}. \textbf{Ordering}, the safety side, asks which enabled transition fires first, whether two are ordered or may interleave, and whether they must fire as one atomic step. \textbf{Eventuality}, the liveness side, asks whether some transition must eventually fire, such as a response sent, a loop exited, or a resource released.

These four components are exhaustive for the model. A run-set is an arrangement of firings, governed by Ordering and Eventuality, over firings whose content is governed by Trigger and Effect, so any under-constraint falls in one of them. They are also independent: a trigger can be missing while its intended effect is clear, ordering can be open while every local transition is fixed, and an eventuality can be open even when every step is individually defined. Real findings often combine components, but the four components name distinct ways a specification can be silent.

\subsection{Gaps, forks, and when they matter}
\label{sec:problem:gapfork}

Gap and fork are not a fifth category but the two forms any of the four components can take, and they call for different remedies. A gap is silence: the specification gives no rule, so the defect is a design omission and the remedy is to add the missing rule. A fork is over-permission: the text does apply, but it is loose enough to admit two or more incompatible readings of the same case, so the remedy is to narrow it until one behavior survives, or to make the choice explicit and negotiated on the wire. Both repairs belong in the specification; they differ in whether a rule must be written or an existing one tightened.
Two conditions decide whether a gap or fork is worth reporting. \textbf{Reachability}: a missing rule in a state no run reaches changes no behavior. \textbf{Observable divergence}: if every admissible reading yields the same outcome the underspecification is latent and we set it aside, whereas readings that accept where another rejects, or emit different responses, are consequential (\S\ref{sec:design:impact}).

\subsection{Challenges}
\label{sec:problem:challenges}

Three aspects of the problem make it hard, and each challenge shapes the design in \S\ref{sec:design}.

\paragraphb{Discovery.} Discovery is hard because the target is an \emph{absence}. Classical audits key on something \emph{present}: a contradiction, an ambiguous term, a transition missing from a model already drawn. An underspecified case offers no keyword and no malformed structure, and does not take shape until one fixes the state and message that expose it. Those pairs are enormous in number, since a procedure's dozens of information elements across several messages and states combine multiplicatively. Enumerating them blindly is infeasible, reading every procedure by hand does not scale, and neither a formal model built up front nor an open-ended request to a language model recovers the missing cases (\S\ref{sec:approach}). Discovery must turn this blind search into a structured one.

\paragraphb{Triage.} A discovered opening does not announce its own remedy. Deciding whether the text is silent about the case or instead covers it too loosely to pin one outcome requires weighing what the cited clauses actually settle, and the answer fixes the repair: write the missing rule, or narrow an over-permissive one.

\paragraphb{Impact and validation.} A finding matters only when its resolutions diverge observably, and establishing that is expensive: candidates are numerous and some protocols lack a second implementation to test against. Assessing impact needs both a conceptual analysis that runs without an implementation and, where implementations exist, execution that confirms the divergence.

\section{Scaffolding with State-Machines}
\label{sec:approach}

Reasoning about underspecification demands two things at once: a precise vocabulary for protocol behavior, and the ability to read the natural-language procedures that define it. A state machine provides the first, and a large language model (LLM) provides the second. Each is necessary, and neither suffices alone.

\subsection{FSMs: right mental model, wrong tool}
\label{sec:approach:fsm}

A communicating state machine is the natural abstraction for a protocol. Protocol behavior is reactive and stateful, and a state machine renders it explicit: each rule is a transition, each state a condition, and the meaning of a message is where the two meet. This is what makes FSM a good basis for reasoning. It is explainable, it localizes a defect to a transition or state, and, as \S\ref{sec:problem} showed, the categories and forms of underspecification fall directly out of it.

Turning a real specification into such a machine, however, is itself the hard problem. Manual formal models pay for their precision with expert effort and narrow scope, a paper-length undertaking per procedure~\cite{basin2018formal, peltonen2021comprehensive}, and recovering a machine from the prose automatically yields models that remain partial and scoped to what could be extracted~\cite{hermes, pacheco2022automated, specgpt}. The difficulty is intrinsic to the documents, which are large, incomplete, and full of implicit state and unstated guards, the very properties that let underspecification exist. No complete and authoritative state machine of these protocols is available to check, so a method that searches for undefined transitions inherits both the cost and the incompleteness of whatever model it starts from~\cite{klischies2023instructions}.

The target also moves. 3GPP revises these specifications at quarterly plenaries and a new release takes two to three years~\cite{3gpp}, so a model faithful to one version soon drifts behind. Full formalization is therefore the wrong primary tool. We keep the state machine for what it is good at, a precise and explainable account of behavior, and use it as a lens and a local instrument rather than a global model to build. As a lens it defines what counts as underspecification and sorts each finding into Trigger, Effect, Ordering, or Eventuality (\S\ref{sec:problem}). As a local instrument it supplies, for one finding at a time, only the small fragment of transitions and candidate completions needed to compare behaviors and derive a test (\S\ref{sec:design}). The entire FSM never has to be built.

\subsection{Na\"ive prompting is not enough}
\label{sec:approach:llm}

What the state machine cannot supply is the reading of the prose that says which transitions exist in the first place. That is where an LLM excels. Protocol procedures are natural-language text threaded with long cross-references, and LLMs are suited to exactly this material: they have been applied directly to specifications and RFCs~\cite{sharma2023prosper, meng2024large}, their context windows hold a procedure together with the clauses it refers to~\cite{IBM_LargerContextWindow_2024}, and their reasoning weighs one reading of a clause against another~\cite{wei2022chain, wei2022emergent, srivastava2023beyond}.

The obvious way to use an LLM is to supply a procedure and ask it to list the underspecifications. \S\ref{sec:eval:rq1} measures how this falls short: against our pipeline on the same procedures, open-ended prompting covers a narrower slice of the protocol, repeats one shallow gap across many information elements, and reaches fewer of the cross-step cases.

The failure is structural, and \S\ref{sec:problem} already named its cause: the target is an absence. Asking a model to list what a document fails to say is an under-constrained generation task, and under-constraint is where LLMs are known to be weak~\cite{wildenburg2024pre}. They collapse onto their most probable output and re-emit one gap template rather than exploring distinct ones~\cite{kirk2024understanding}; they default to local, single-step completion and under-explore the reasoning about arrangement and change across a run that Ordering and Eventuality require~\cite{kambhampati2024can, valmeekam2023planbench}. Asserting that a specification does not constrain some case is a claim of absence over a document, which a model trained to guess, not to abstain, will make whether or not it holds~\cite{kalai2025why, truong2023language}. A common refinement, iterating the prompt with its own best findings as few-shot examples for the next round, does not supply the missing breadth either. Few-shot examples steer a model's output distribution toward themselves~\cite{zhao2021calibrate}; so seeding later rounds with prior findings keeps the search near those prototypes and rediscovers the same gap families instead of surfacing new ones (\S\ref{sec:eval:rq1}).

\S\ref{sec:intro} reported that model capability does not make the scaffold useless. When we expanded capability inside one family from gpt-5.4 nano to mini to full (isolating the workflow from the model), we found that cases spanning several protocol steps (for Ordering and Eventuality) are ones that a local prompt misses most: our pipeline on the weakest model finds more of them than naïve prompting finds on the strongest, roughly 1/2 against 1/3, a gap that never closes (\S\ref{sec:eval:rq1}). Removing the scaffold entirely by letting two frontier agentic models direct their own process over whole specifications also missed several underspecifications. What the scaffold contributes is therefore not knowledge the model lacks but a standard of admission it has no reason to apply on its own, a property of the problem rather than of any model generation.

\subsection{Putting the two together}
\label{sec:approach:together}

The two limitations are complementary, which we can exploit. A state machine has the structure that prompting lacks but cannot read the prose; an LLM reads the prose but, if unconstrained, lacks the structure. \sys places the state machine in front of the LLM as a reasoning scaffold rather than a model to construct: the state-machine view turns an open-ended request into bounded, typed, and located questions over named transitions, and the LLM answers each one from the specification text. Decomposing an ill-posed task into constrained sub-questions is the established way to steer a model clear of these failure modes~\cite{zhou2023least}. The scaffold supplies structure, process, and coverage; the LLM supplies the reading. The pipeline of \S\ref{sec:design} is how the two are combined.

The division also says which half is affected by model capability. The reading improves with every model generation, and \sys inherits that improvement unchanged. What the scaffold holds fixed, a lens for what counts as an opening and a deterministic verdict on whether it matters, answers the question instead of the model.

\section{Design of \sys}
\label{sec:design}

\begin{figure*}[t]
  \centering
  \includegraphics[width=0.85\textwidth]{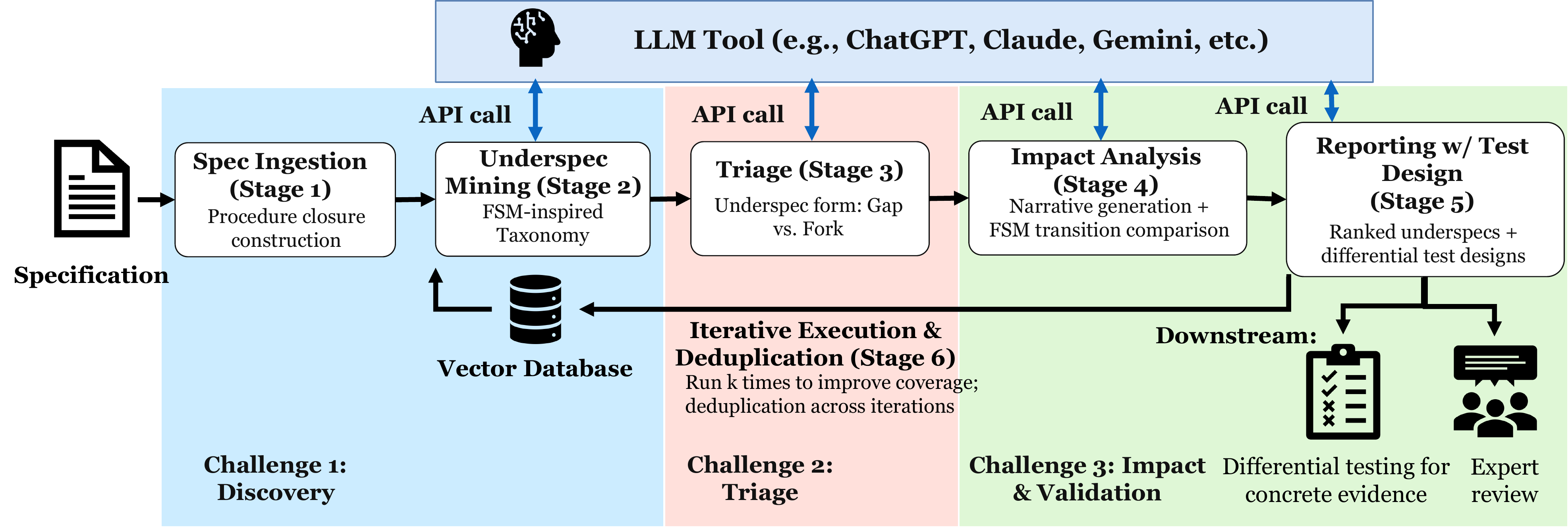}
  \vspace{-0.05in}
  \caption{The \sys pipeline.}
  \label{fig:pipeline}
\end{figure*}

\S\ref{sec:approach} argued for placing the state machine in front of the LLM as a scaffold. \sys realizes that scaffold as the six-stage pipeline of Figure~\ref{fig:pipeline}. Each stage poses the LLM one bounded question, grounded in specification text and answered as typed JSON, while the scaffold owns everything around it: the state passed between stages, the iteration that drives coverage, and the judgments that decide whether a finding matters.

Two principles shape the design. The task is decomposed using the CFSM of \S\ref{sec:problem}. Each stage asks about one part of the model: a transition's trigger or effect, the arrangement of firings over a run, the form of an opening, or the completions that fill it. The consequential judgments are computed by a deterministic FSM engine rather than by the LLM: the model extracts a small state machine from the prose; the engine decides which behaviors diverge and how to tell them apart. Appendix~\ref{app:prompts} gives the prompt for each stage's question.

The stages map to the three challenges of \S\ref{sec:problem:challenges}. Stages~1 and~2 address discovery, Stage~3 triage, and Stages~4 and~5 impact and validation, while Stage~6 runs underneath, iterating until coverage saturates. Popping up a level, the pipeline instantiates a recipe that is not specific to protocols: choose a lens, turn it into bounded typed questions, and adjudicate deterministically. This is the core of our scheme; by contrast, the closure construction and deduplication thresholds are engineering details other implementations could replace.

\subsection{Stage 1: Context construction}
\label{sec:design:ingest}

3GPP and O-RAN specifications are organized around \emph{procedures}: named message exchanges that each accomplish one task, such as UE Context Release or E2 Setup, with a dedicated subsection listing a procedure's messages, information elements, and handling rules. Implementers reason at this granularity, and so does \sys: its unit of analysis is the procedure, not a fixed-size document chunk. A procedure's full meaning is nonetheless scattered across non-contiguous sections, message and IE tables, so Stage~1 builds a \emph{procedure closure}: it locates the sections whose titles or text match the target procedure, renders their message and IE tables as Markdown, and resolves local clause references recursively, inlining each referenced clause and guarding against reference cycles. The result is a self-contained bundle carrying the same cross-reference closure a human reader would assemble. When it is still insufficient, the mining stage may request additional sections once from a table of contents, under a fixed character budget, rather than silently reasoning from a partial view.

\subsection{Stage 2: Taxonomy-based mining}
\label{sec:design:mining}

Mining turns the four categories of \S\ref{sec:problem:categories} into the search itself, in two scoped passes over the closure. A \emph{local} pass mines Trigger and Effect, the content of a single transition, and a \emph{temporal} pass mines Ordering and Eventuality, the arrangement of firings across a run. Splitting them bounds each prompt to a single category.

The \emph{temporal} pass is gated on applicability, because concurrency and loops are the structural sources of temporal underspecification, and most procedures have neither. Ordering is mined only when the procedure admits concurrent or overlapping transactions over shared state. Eventuality is mined only when a run can fail to progress through a loop, a retry, or a pending obligation that is not forced to occur. A strictly sequential request and response admits neither, and the pass returns nothing rather than manufacture a finding. Across both passes, every finding must cite the messages, IEs, and sections it concerns and must argue reachability with a concrete run. Breadth comes from examining every transition and message using these two passes instead of an open-ended request, which is why the search does not collapse onto one template (\S\ref{sec:eval:rq1}).

\subsection{Stage 3: Gap/fork triage}
\label{sec:design:triage}

Stage~3 assigns each finding its form. Reading the cited excerpts, the LLM decides whether the specification pins a single required behavior for the scenario: silence makes the finding a \textbf{Gap}, and text loose enough to admit two or more incompatible behaviors makes it a \textbf{Fork}, the $|\Delta|=0$ versus $|\Delta|\ge 2$ distinction of \S\ref{sec:problem:gapfork} applied to one finding. Whether the looseness was intended is a separate question the model cannot settle, so the stage records its reading without acting on it, and it defaults to Gap when the evidence is unclear.

\subsection{Stage 4: Impact analysis}
\label{sec:design:impact}

Stage~4 decides whether a finding is consequential, and it is where the deterministic scaffold does the most work. It runs in two steps whose separation is the point. First, the LLM enumerates two to four permissible \emph{completions}, distinct behaviors the cited text allows, varied by category: competing trigger rules, differing state updates, orderings, or liveness assumptions for eventuality. It then extracts a small partial CFSM fragment from the spec text: the transitions the specification does pin down, the exposing configuration with a concrete reachability trace, the under-constrained slot, and each completion rendered as a transition with a decision, one of accept, reject, ignore, or no-progress, and a progress flag. When the configuration or the slot cannot be grounded in the text, the model is instructed to leave those fields empty.

\begin{table}[t]
  \centering
  \setlength{\tabcolsep}{4pt}
  \begin{tabular}{@{}p{0.45\columnwidth}ccc@{}}
    \toprule
    \textbf{Impact (severe to benign)} & \textbf{Decision} & \textbf{State} & \textbf{Output} \\
    \midrule
    Conflicting Actions          & \xmark &        &        \\
    Inconsistent Internal States & \cmark & \xmark &        \\
    Inconsistent Notifications   & \cmark & \cmark & \xmark \\
    Latent (dropped)             & \cmark & \cmark & \cmark \\
    \bottomrule
  \end{tabular}
  \caption{The impact ladder as a cascade over three dimensions: the decision, the resulting state, and the emitted output. \cmark\ marks a dimension the pair agrees on and \xmark\ one it differs on.}
  \label{tab:impact-ladder}
\end{table}

Second, a deterministic engine, not the LLM, computes the verdict. A finding whose reachability trace or slot is missing is flagged as un-modelable and set aside, so no severity is invented. The engine then compares every pair of completions along three dimensions in decreasing severity, its decision, its resulting state, then its output, as in Table~\ref{tab:impact-ladder}, and the first dimension on which the pair differs fixes its class. Only the state comparison needs more than a syntactic check: two different next states count as a divergence only if some later message would treat them differently. Pairs that agree on all three are latent and dropped, and a findings worst surviving class prioritizes it for review and testing. Deciding severity mechanically keeps the LLM from grading its own findings, and the distinguishing message the state check finds becomes the discriminator the generated test probes with.

\subsection{Stage 5: Test design}
\label{sec:design:test}

Stage~5 turns a consequential finding into a differential test, the clearest instance of the design's central split. \sys first assembles a test plan mechanically from Stage~4's artifacts: the setup is the reachability trace that drives an implementation to the exposing configuration, the trigger is the pending message; each completion becomes one branch carrying its decision, next state, and expected outputs. Each impact level yields a discriminator, observing the accept-or-reject decision, probing with the engine's distinguishing input, comparing emitted notifications, or watching whether progress occurs. Only then is the LLM invoked, for concretization alone: it maps the symbolic states and messages to real message and IE names and fills mandatory IEs with realistic values, but may not change the branch expectations or the discrimination method. The test's logic is therefore fixed by the model and traceable to it, while the LLM supplies the protocol-specific values it is best equipped to know.

\subsection{Stage 6: Iteration and deduplication}
\label{sec:design:dedup}

A single mining pass is probabilistic and does not always cover a complex procedure, so \sys iterates. Each iteration re-runs mining from a fresh start and deliberately does \emph{not} seed the model with earlier findings, since showing the model its own prototypes narrows the search rather than broadening it (\S\ref{sec:approach}). A two-stage filter deduplicates each run's output against a persistent vector store: a keyword filter proposes candidates sharing the finding's category and overlapping in its messages, IEs, or sections, and an embedding check removes any whose scenario description is within a cosine similarity threshold of an existing one. Novel findings are added to the store for the next iteration. Mining stops when two consecutive runs are at least eighty percent duplicates. This is what makes the tool terminate and bounds its cost to a few passes per procedure (\S\ref{sec:eval:rq4}).

The six stages produce a ranked report: for each underspecification, its category and form, the competing interpretations, the predicted impact, and a differential test design. That single output supports two independent downstream uses. A standards reviewer can audit whether the proposal is genuinely unresolved by the normative text, which needs no implementation and applies even to protocols for which none exists. Where two comparable implementations do exist, differential testing confirms that they resolve the opening differently. The two establish different things, standards-level plausibility and real implementation consequence, and \S\ref{sec:eval:rq2} and~\S\ref{sec:eval:rq3} evaluate each in turn.

\section{Implementation}
\label{sec:impl}

\sys owns the workflow and treats the LLM as a replaceable component; a stronger backend improves it with no pipeline changes. The system does the durable work, constructing the context, tracking progress and validating each stage's output, while the LLM contributes only targeted reasoning at each stage. The implementation is about 5{,}500 lines of Python, of which two components carry the weight of the design: the deterministic FSM engine, which performs the reasoning the pipeline does not delegate to an LLM, and the component-test harness, which confirms a finding against real implementations. We plan to publicly release the code, the prompts, and the test harness.

\subsection{The deterministic FSM engine}
\label{sec:impl:engine}

The engine is about 575 lines. It consumes the partial CFSM fragment that the model-extraction stage emits and returns an impact verdict and a symbolic test plan, in three steps. \emph{Loading} parses the JSON into typed structures, coercing tolerantly so that a model that emits a bare string for a list, or omits a field, is normalized or flagged rather than let it crash the run. \emph{Judging} applies the ladder of \S\ref{sec:design:impact}; the one step beyond comparing recorded fields is a small bounded search over the specified transitions that decides whether two resulting states are observably distinct. \emph{Planning} turns the verdict, its branches, and the distinguishing message into the symbolic skeleton the concretization stage fills in. Every step except the model's own extraction is deterministic, so the verdict is reproducible.

\subsection{Provider-agnostic pipeline}
\label{sec:impl:pipeline}

The scaffold around the engine is model-agnostic, calling the LLM only for targeted reasoning at each stage, so every backend we test runs the full pipeline unchanged (\S\ref{sec:eval:rq4}). \textsf{python-docx} parses the Word specifications into a section tree from which Stage~1 assembles the procedure closure. A unified client interface abstracts the provider's APIs, and each stage's output is validated against the expected structure, with tolerant coercion rather than a rigid grammar, before the next stage consumes it. A persistent \textsf{chromadb} vector store holds accepted findings, with embeddings fixed to one provider so a finding's identity stays stable no matter which backend mined it.

\subsection{Component-test harness}
\label{sec:impl:harness}

Validating a finding (\S\ref{sec:eval:rq3}) means checking whether two independent implementations resolve the underspecification differently on production code. The harness compiles the generated test against the pinned sources of OpenAirInterface~\cite{oai5g} with its bundled FlexRIC~\cite{flexric}, and srsRAN~\cite{srsran}. It then invokes the relevant handler in the tested process, and stubs the surrounding SCTP, MAC, RRC, and DU-manager dependencies, so an observed divergence is attributed to control-plane logic rather than to transport or timing. Cases that end in a process abort run in a forked subprocess, and liveness cases advance a manual clock instead of waiting, so every test is deterministic and quick. All results are pinned to fixed codebase commits.\footnote{OpenAirInterface \texttt{09c9bb99}, its bundled FlexRIC \texttt{ef6d722f}, and srsRAN \texttt{4bf15439}.} A smaller subset of findings is additionally confirmed end-to-end on a live OAI and srsRAN testbed (Appendix~\ref{app:e2e}).

\section{Evaluation}
\label{sec:eval}

\sys produces, for each procedure, a ranked set of underspecifications, each with its interpretations, its predicted impact, and a differential test. We answer four questions:

\textbf{Effectiveness} (\S\ref{sec:eval:rq1}): Does the FSM-scaffolded pipeline discover underspecifications better than naïve prompting?

\textbf{Proposal validity} (\S\ref{sec:eval:rq2}): Do domain experts find the proposals grounded in the standard?

\textbf{Downstream utility} (\S\ref{sec:eval:rq3}): Do the proposals expose real divergence between implementations?

\textbf{Feasibility} (\S\ref{sec:eval:rq4}): Is the workflow affordable, terminating, and portable across model backends.

This evaluation leaves two questions open: it reports no recall, since the space of valid-but-unhandled cases admits no ground truth, and it scopes each analysis to a single procedure, leaving cross-procedure interactions for future work.

\subsection{Setup}
\label{sec:eval:setup}

We run \sys on the 36 procedures that OpenAirInterface and srsRAN both implement, drawn from four RAN control-plane protocols (F1AP, NGAP, E1AP, and E2AP), the GTP-U user plane, and the E2SM-KPM service model; Appendix~\ref{app:procedures} lists them. We use three backends: \textbf{gpt-5.4-mini}~\cite{gpt54mini}, \textbf{claude-haiku-4.5}~\cite{claudehaiku45}, and \textbf{gemini-3-flash}~\cite{gemini3flash}.

The four studies use different populations, which we state once here. Effectiveness pools the pipeline's findings and a naïve prompt's findings across all three backends, and adds two smaller comparisons that vary the model rather than the method: a capability sweep inside one model family, and an agentic setting in which two frontier models mine whole specifications while directing their own process and tool use. Proposal validity uses the 197 proposals that gpt-5.4-mini delivers; of these, 40 are confirmed by static review of the two implementations, and the remaining 157 are audited against the specification text by experts. Downstream utility tests every delivered finding, from any LLM backend, whose scenario the two implementations actually exercise; 60 of these confirm a divergence (\S\ref{sec:eval:rq3}). Feasibility compares the three backends head to head.

Two of the studies rely on judgments rather than execution. Proposal validity is decided by two standards-literate experts. Effectiveness is scored by a deterministic checklist computed from each finding's text and by two blinded LLM judges from different model families (GPT-5.4 and DeepSeek-v4-Pro), which decide only concrete, checkable properties rather than quality (Appendix~\ref{app:rq1}). Because the two judges and the checklist agree, no conclusion rests on a single model's opinion. Runtime divergence, in contrast, is decided objectively by the compiled component tests of \S\ref{sec:impl:harness}.

\subsection{Effectiveness against naïve prompting}
\label{sec:eval:rq1}

\begin{table}[t]
  \centering
  \setlength{\tabcolsep}{5pt}
  \begin{tabular}{@{}lcc@{}}
    \toprule
    & \textbf{\sys} & \textbf{Naïve} \\
    \midrule
    \multicolumn{3}{@{}l}{\emph{Coverage and redundancy (deterministic)}} \\
    Distinct sections touched          & \textbf{336} & 287 \\
    Distinct messages touched          & \textbf{127} & 103 \\
    Repeated missing-rule findings & \textbf{155} & 269 \\
    \midrule
    \multicolumn{3}{@{}l}{\emph{Grounding and reach (blinded judge)}} \\
    Cite two or more sections          & \textbf{83\%} & 66\% \\
    Reviewer-ready                     & \textbf{88\%} & 68\% \\
    Cross-step (temporal) finding      & \textbf{58\%} & 42\% \\
    \bottomrule
  \end{tabular}
  \caption{Discovery quality on 525 \sys findings and 551 naïve-prompting findings, pooled over the three backends. Judge rows are the primary judge; the second judge agrees (\S\ref{sec:eval:rq1}).}
  \label{tab:rq1}
\end{table}

The baseline is the obvious way to use an LLM: the same procedure closure, a direct instruction to list the underspecifications, and none of the taxonomy of Stage 2 (Figure~\ref{fig:prompt-naive}). We compare 525 findings from the \sys pipeline against 551 from naïve prompting, produced under all three backends and judged blind (\S\ref{sec:eval:setup}). To keep the comparison fair, the pipeline findings are scored on their bare claim, the same fields the baseline exposes. Table~\ref{tab:rq1} collects the headline results; Appendix~\ref{app:rq1} defines each instrument.

The pipeline covers more of each protocol. It touches 336 distinct specification sections and 127 distinct messages, against 287 and 103 for the baseline, and it does so while repeating itself less. Clustering findings by the protocol locus they name, 269 of the naïve findings only restate a missing-rule pattern already recorded at the same location, against 155 for the pipeline; the taxonomy keeps moving the model to a fresh transition, whereas an open prompt falls back to one high-probability template, ``optional IE absent, so behavior is undefined,'' instantiated per field.

The judged metrics agree.\footnote{Every reviewer-ready and cross-step comparison reported in this subsection is significant at $p<10^{-6}$ by Fisher's exact test.} The pipeline grounds each claim more densely, with 83\% of findings citing two or more sections against 66\%, and a larger share is reviewer-ready, meeting at least four of six concreteness criteria (Appendix~\ref{app:rq1}): 88\% against 68\% under the first judge and 86\% against 67\% under the second. It also reaches more of the ordering and eventuality cases a local prompt tends to skip: 58\% against 42\% under the first judge and 54\% against 39\% under the second.

Refining the baseline does not close the gap, as \S\ref{sec:approach:llm} predicted. Feeding a round's well-grounded findings back as few-shot examples steers the next round toward those prototypes, so the rounds converge on the same issue families: seed similarity rises (0.67 against 0.64 for an unseeded control) while within-round diversity falls (0.41 against 0.44).\footnote{Appendix~\ref{app:rq1} gives the metrics.} Refinement constrains the answer space, while the taxonomy shows the whole problem space.

The pipeline and the baseline use the same backend, so the pipeline's advantage cannot come from a stronger model; to confirm it comes from the workflow, we do a capability sweep within one family, from gpt-5.4 nano to mini to full, and run both at every tier (Table~\ref{tab:rq1-prog}). The pipeline wins on breadth, redundancy, and cross-step reach at every tier, and its run on the weakest model beats naïve prompting's run on the strongest: 50 to 60\% of its findings using gpt-5.4-nano span several protocol steps, against about 30\% for naïve prompting using gpt-5.4-full. A stronger model does let naïve prompting ground itself, which closes the grounding gap. However, it never explores the temporal cases sufficiently: its share holds near 30\% from the smallest model to the largest, while the pipeline stays near 50\% regardless of model. The gains are due to the workflow; a stronger backend lifts both without letting the naïve prompt catch up with the cases the taxonomy targets.

\begin{table}[t]
  \centering
  \setlength{\tabcolsep}{6pt}
  \begin{tabular}{@{}lcc@{}}
    \toprule
    \textbf{Primary category} & \textbf{Findings} & \textbf{Share} \\
    \midrule
    Genuine underspecification      & \textbf{244} & 46\% \\
    \midrule
    Optionality read as a gap       & 96 & 18\% \\
    Deliberate policy freedom       & 89 & 17\% \\
    Malformed input, not valid syntax & 38 & 7\% \\
    Delegated to another document   & 30 & 6\% \\
    Editorial or reference defect   & 27 & 5\% \\
    Duplicate finding               & 6  & 1\% \\
    \bottomrule
  \end{tabular}
  \caption{One primary category per finding for the 530 that gpt-5.6-sol and claude-opus-5 mined agentically over the same six protocols.}
  \label{tab:noharness}
\end{table}

A second natural question is how well agentic mining does with no scaffold at all. We give the naïve prompt of Figure~\ref{fig:prompt-naive}, unchanged, to two frontier models, gpt-5.6-sol~\cite{gpt56sol} and claude-opus-5~\cite{claudeopus5}, with an entire specification as context and no imposed workflow: each decides how to proceed, writing scripts and calling tools. The 530 findings this produces broaden the search and ground themselves about as densely as the pipeline does, so capability substitutes for part of what the workflow supplies, but not for the standard of what counts. Assigning each finding one primary category against the cited text (Table~\ref{tab:noharness}), 244 are genuine underspecifications using the definition of \S\ref{sec:problem}, while 185 treat freedom the standard grants on purpose as a failure to specify: e.g., one calls QoS pre-emption underspecified while quoting the clause that makes the process operator-dependent.

Beyond precision, agentic mining does poorly on recall. In a four-procedure case study cross-checking 36 pipeline findings against the same models' output, claude-opus-5 recovers 14 completely, while 22 are at best partial, naming the feature family but not the transition that makes a claim reviewable. Capability thus broadens untriaged mining without disciplining or sharpening it. Appendix~\ref{app:rq1} details both the breakdown of cross-checks and the missed findings.

These are the gains the decomposition predicts (\S\ref{sec:approach}). Turning one open-ended request into bounded, located questions makes the search cover more transitions, repeat itself less, and reach the temporal cases. An open-ended prompt results in local completion. While a stronger model can match the grounding of our workflow, it cannot emulate our pipeline's breadth and temporal reach.

\subsection{Proposal validity}
\label{sec:eval:rq2}

\begin{table}[t]
  \centering
  \setlength{\tabcolsep}{6pt}
  \begin{tabular}{@{}lccc@{}}
    \toprule
    & \textbf{Gap} & \textbf{Fork} & \textbf{Total} \\
    \midrule
    Effect       & 24 & 71  & \textbf{95} \\
    Trigger      & 13 & 15  & 28 \\
    Ordering     & 0  & 17  & 17 \\
    Eventuality  & 3  & 2   & 5 \\
    \midrule
    \textbf{Total} & 40 & 105 & \textbf{145} \\
    \bottomrule
  \end{tabular}
  \caption{The 145 unanimously accepted proposals by category and form.}
  \label{tab:rq2-repair}
\end{table}

A proposal is valid if the underspecification it raises is genuinely unresolved by the normative text, not hallucinated. We check the 197 proposals gpt-5.4-mini delivers in two ways. For 40 whose scenario both implementations exercise, static review confirms that they resolve it differently. The remaining 157 were examined by two standards-literate experts, who read each proposal against the cited specification and marked it as an underspecification when no on-point statement in the specification settles it.

The experts accept the vast majority. Individually they accept 94.3\% and 96.8\% of the 157 proposals; together they accept 145 unanimously (92.4\%) and 155 with at least one vote (98.7\%), and flag only 2 as false positives in common, agreeing on 147 of the 157 (Appendix~\ref{app:rq2}). Folding in the 40 confirmed by code-review, 185 of the 197 proposals are conservatively supported, accepted by both experts or code-review, and 195 are supported by code or at least one expert, so the proposals are well-grounded.

Because the accepted proposals carry a category and a form, they also indicate what repairing underspecification requires (Table~\ref{tab:rq2-repair}). First, forks outnumber gaps roughly two to one. Second, regardless of form, the under-constrained part is usually the effect (the resulting state or output) not the trigger or the ordering. Since the effect accounts for 95 of the 145 proposals, stating the post-state, output, and cleanup alone would address nearly two-thirds of underspecifications. The fixes fall into a few recurring patterns, one per category and form. For instance, an effect gap is closed by adding a postcondition, a trigger gap by completing the receive table. Appendix~\ref{app:rq2} lists each underspecification pattern together with a draft repair template.

\subsection{Downstream utility: differential testing}
\label{sec:eval:rq3}

\begin{table}[t]
  \centering
  \setlength{\tabcolsep}{3pt}
  \begin{tabular}{@{}lccccccc@{}}
    \toprule
    & NG & F1 & E1 & E2 & GTP-U & E2SM & \textbf{Total} \\
    \midrule
    Validated cases & 21 & 19 & 9 & 7 & 2 & 2 & \textbf{60} \\
    \bottomrule
  \end{tabular}
  \caption{The 60 differential-test-validated divergences by protocol.}
  \label{tab:rq3-protocol}
\end{table}

A proposal has downstream utility if two independently built implementations resolve the underspecification differently once the test runs. We compile every underspecification that both implementations exercise (from any LLM backend) against the pinned OAI and srsRAN sources, and run it as a software component test (\S\ref{sec:impl:harness}).

60 grouped cases produce a real divergence, spread across all six protocols (Table~\ref{tab:rq3-protocol}; a full list is in Appendix~\ref{app:divergences}). By the observed outcome, 32 are conflicting actions such as accept versus reject; 16 are inconsistent internal states, exposed only by a later operation; and 12 are inconsistent notifications. The findings are therefore not merely textual, and a deployment that adds more implementations can only exacerbate these divergences. The generated test designs carry most of the weight, reducing the effort of writing manual tests: of the 60, 37 run exactly as designed and 19 more keep the generated trigger while adapting only the harness that observes it for implementation specific reasons.

Stage~4 also predicts each finding's impact class from the specification text alone, before any code runs. The prediction is deliberately conservative, highlighting the worst outcome the text permits while two implementations often converge on a milder one, so an exact match understates it: the worst-outcome prediction matches 63\% of the validated cases exactly, and the observed outcome falls within the engine's full predicted set for 72\%. Since the prediction only ranks which findings to test first, highlighting potential severity is crucial.

The divergences recur along one axis: OAI tends to tolerate a stale item and proceed, to lack a liveness timer, or to reclaim state late. On the other hand, srsRAN validates up front, rejects, and bounds its waits. Several legal messages reach an assertion that aborts the OAI or FlexRIC process. The categories of \S\ref{sec:problem:categories} are thus not protocol-specific. We now describe two cases concretely; Appendix~\ref{app:cases} provides more.

\paragraphb{Releasing one session versus dropping the connection.} A \textsf{PDU SESSION RESOURCE RELEASE COMMAND} lets the core network tell a base station to free some of a device's data sessions, each named by a \textsf{PDU Session ID}. The specification gives no rule for a \textsf{PDU Session ID} that names a session which is not currently active, a stale or duplicated entry that a live network produces routinely. OAI skips the stale ID and frees the sessions that remain, keeping the device attached; srsRAN rejects the whole command and tears down the device's entire context, every session and the signaling connection with it. The same message thus frees one session on OAI and disconnects the device on srsRAN, and an attacker able to inject a single bogus \textsf{PDU Session ID} can turn a routine release into a full disconnection.

\paragraphb{Keeping a conflicting identity versus reconciling it.} A \textsf{DOWNLINK NAS TRANSPORT} message names one signaling association by a pair of identifiers. The specification gives no rule for a message whose \textsf{AMF-UE-NGAP-ID} is already bound to a different \textsf{RAN-UE-NGAP-ID}, a stale or spoofed pairing a live network can produce. OAI keys on the \textsf{RAN-UE-NGAP-ID} alone and leaves both pairings live; srsRAN reconciles them and returns an Error Indication. On OAI the surviving pairing is a context-confusion primitive: an attacker who injects one crafted pair can leave it standing, so later signaling meant for that device can be delivered to the wrong device.

\subsection{Operational feasibility}
\label{sec:eval:rq4}

\begin{table}[t]
  \centering
  \setlength{\tabcolsep}{4pt}
  \begin{tabular}{@{}ccccc@{}}
    \toprule
    \textbf{Backend} & \textbf{cost (\$)} & \textbf{time (min)} & \textbf{saturated} & \textbf{runs/proc.} \\
    \midrule
    gpt-5.4-mini     & 0.38 & 1.84 & 36/36 & 4.7 \\
    claude-haiku-4.5 & 0.21 & 0.52 & 36/36 & 3.3 \\
    gemini-3-flash   & 0.08 & 0.78 & 36/36 & 4.1 \\
    \bottomrule
  \end{tabular}
  \caption{Cost, latency, and termination per backend. Cost and latency are per delivered proposal, across the whole pipeline.}
  \label{tab:rq4}
\end{table}

To be usable at scale the workflow must be cheap, must terminate, and must run on whatever backend is at hand. All three backends complete the full pipeline unchanged, which settles portability; Table~\ref{tab:rq4} reports the rest. A delivered proposal costs cents, not dollars, from \$0.08 on gemini to \$0.38 on gpt-5.4-mini across the whole pipeline, and takes between half a minute and two minutes of wall time, the spread reflecting how much reasoning each model is configured to do.

\begin{figure}[t]
  \centering
  \includegraphics[width=\columnwidth]{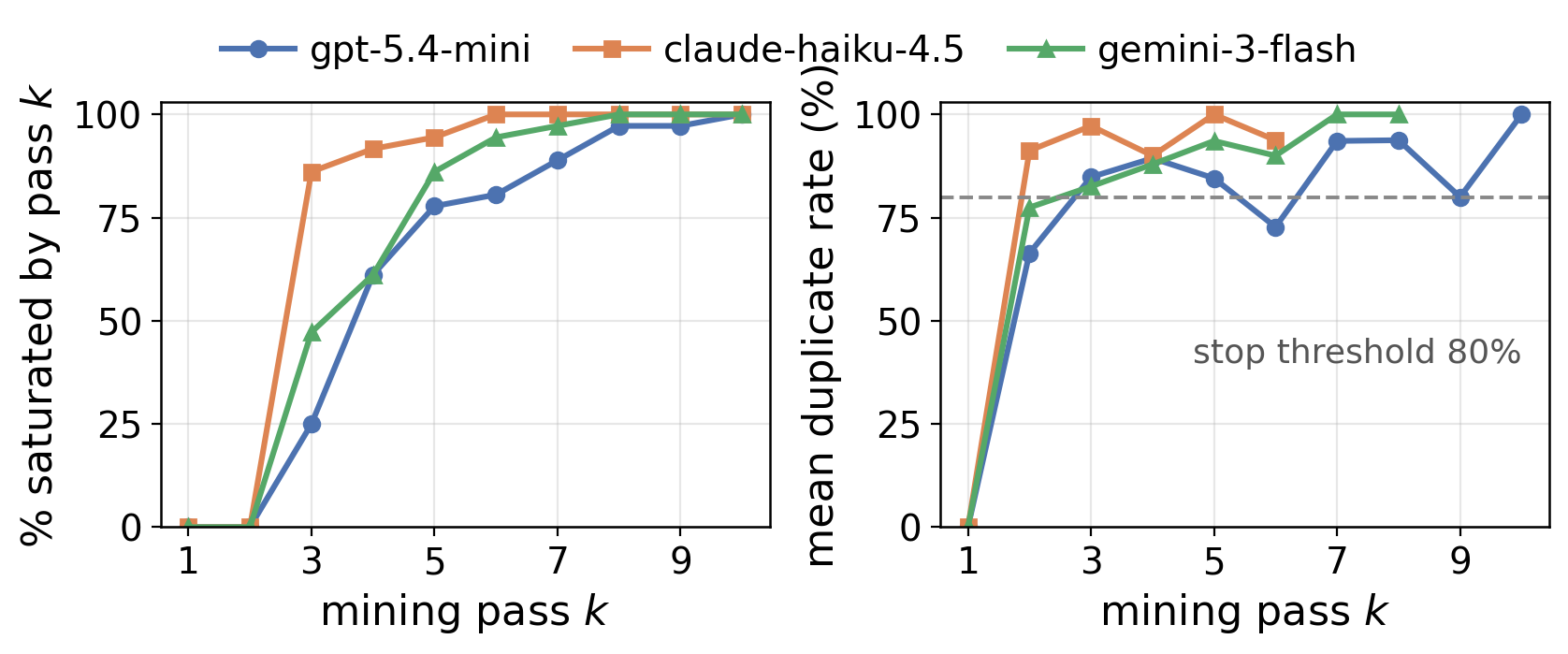}
  \vspace{-0.05in}
  \caption{Mining saturates within a few passes. \iffalse Left: the share of procedures that have finished rises steeply with the pass count. Right: the duplicate rate crosses the 80\% stop threshold by the second or third pass. \fi}
  \label{fig:saturation}
\end{figure}

Mining runs are also bounded (Figure~\ref{fig:saturation}). All 36 saturate for all three backends, in 3 to 5 passes each, so a procedure costs a small bounded multiple of one mining pass rather than an open-ended budget. Combined with the per-proposal cost, a full sweep of the six protocols costs tens of dollars.

The backends do not, however, find the same things. Matched strictly by the deduplicator of \S\ref{sec:design:dedup} or leniently by a manual conceptual pass, only 5 to 17\% of gaps are found by all three, and 28 to 54\% by at least two. How to combine the backends is therefore strategic: their union maximizes coverage for a recall-oriented pre-release audit, their intersection maximizes precision for a fast high-confidence screen, and a single backend is the economical default for routine use.

\section{Generalizing to DNS}
\label{sec:dns}

Underspecification is a property of any standard that pairs a formal message syntax with natural-language procedures, not a feature of 5G, so we test it on the DNS RFCs. Porting \sys replaced under 10\% of the workflow: the six stages ran unchanged, and the specification parser, prompt framing, and additive cross-RFC reference graph were the only substantive changes (Appendix~\ref{app:dns}). 
We mine a handful of procedures across the RFC family and confirm each finding differentially on the Ferret testbed~\cite{kakarla2022scale} against independent authoritative servers: BIND~\cite{bind9}, Knot~\cite{knotdns}, Technitium~\cite{technitium}, YADIFA~\cite{yadifa}, and Hickory DNS~\cite{hickorydns}.

The pipeline surfaces genuine DNS underspecifications of both forms of the taxonomy. We present one effect fork here and give the rest in Appendix~\ref{app:dns}. RFC~2136~\cite{rfc2136} requires an accepted update to advance the zone's SOA serial number but does not fix the increment or 
granularity. RFC~1982~\cite{rfc1982} allows any positive delta. Hence, one three-mutation update advanced the number by one on BIND and Knot and by three on Technitium. This carries less operational weight than the 5G findings, yet each is a genuine ambiguity the text leaves open. Note, however, that a DNS contract is spread across a family of RFCs, so establishing that a gap is genuinely open requires confirming that no sibling RFC closes it.

\section{Related Work}
\label{sec:related}

\paragraphb{Testing implementations against specifications.} A large body of work checks an implementation against its specification. Conformance testing derives cases from the standard~\cite{contester, netestllm, ipanda, friden20205g, zhang2019survey}, protocol fuzzing perturbs a live stack~\cite{llmif, meng2024large, chatfume, lipfuzzer, mgptfuzz}, model-based testing generates suites from a behavioral model~\cite{eywa}, and differential and parser-validation methods compare independent implementations~\cite{parval}. Closest to our domain, several frameworks test 5G and O-RAN interfaces and cores directly~\cite{tan2025automated, bennett2024ransacked, yang2024oranalyst, khichane20245gc}. These methods share one assumption: the specification is the oracle of correct behavior, so a divergence is the implementation's fault. Studies of the specifications themselves show that assumption is not always safe~\cite{raza2019systematic, cellularlint}. \sys works upstream, asking whether the specification determines a behavior at all and running differential tests to confirm a gap rather than to grade an implementation.

\paragraphb{Formalizing and analyzing specifications.} A second line of work builds formal models to verify protocol behavior, either constructing state machines by hand~\cite{basin2018formal, zhang2019formal, peltonen2021comprehensive, hou2021discovering} or extracting them from prose with NLP and, more recently, LLMs~\cite{hermes, pacheco2022automated, specgpt, flowfsm, yuan2023ambiguity, chen2021bookworm}. Our approach descends from the same lineage, communicating finite-state machines~\cite{brand1983cfsm} and the safety and liveness decomposition of a run~\cite{alpern1985defining}, but put to a different use. Model checking over such a model can flag an undefined transition~\cite{klischies2023instructions}, yet the finding is only as sound as the model, and building a faithful whole-protocol model of a large, frequently revised specification is itself the bottleneck (\S\ref{sec:approach}). \sys never builds one: it reads the partial state machine the specification already induces, one finding at a time. An LLM does the reading, but reading what a specification says is not finding what it fails to say, a task on which post-trained models are documented to do less well~\cite{kirk2024understanding, kambhampati2024can, valmeekam2023planbench, kalai2025why, truong2023language}.

\paragraphb{Ambiguity and underspecification detection.} Closest to our paper is work that looks for imprecision in a specification directly, though the term ``ambiguity'' is overloaded and covers several distinct concepts (Figure~\ref{fig:underspec_comparison}). A sentence may admit more than one grammatical parse (syntactic)~\cite{sage}; a transition may be missing from a hand-built state machine (structural)~\cite{klischies2023instructions}; a word may carry more than one meaning (lexical)~\cite{mahbub2024can, wildenburg2024pre, bansal2025can}; two clauses may contradict or a reference may dangle (logical)~\cite{cellularlint, pawagi2025rfcscope}; a request may be open to more than one reading of intent~\cite{mondal2025tackling}; and analogous benchmarks target legal text~\cite{hendrycks2021cuad, koreeda2021contractnli, choudhury2025better}. All such earlier work targets something present, a token or structure with a signal to key on. Semantic underspecification is the absence beneath the text, the level prior work has largely ignored.

\section{Open Questions}
\label{sec:discussion}

\paragraphb{Scope:} \sys targets standards that pair a formal message syntax with natural-language procedures, not deployment-oriented documents such as GSMA operational specifications~\cite{gsma}. It locates where the text leaves behavior open (\S\ref{sec:problem:gapfork}); automating the fix is an open problem. Because the expert-audit track validates findings against the text alone (\S\ref{sec:eval:rq2}), \sys can act as a pre-release quality gate for standards bodies.

\paragraphb{From interoperability to security:} We have framed an opening as a vendor-interoperability hazard, but several validated findings are already exploitable: one crafted identifier pair plants a context-confusion primitive (\S\ref{sec:eval:rq3}); one legal message reaches an assertion that ends a base-station process, and an under-constrained measurement filter can return every device a base station tracks to a subscription meant to scope one slice (Appendix~\ref{app:cases}). What changes under a security framing is the criterion: an adversary asks which openings a \emph{sender} can exploit from outside, not which ones two receivers resolve differently. That is a different search over the same taxonomy; we only ran the second one.

\paragraphb{Across not within specifications:} \sys reasons over a procedure closure inside a single document (\S\ref{sec:design:ingest}), and that boundary is a choice we should relinquish. DNS forced the first step, since a DNS contract is spread over a family of RFCs; establishing that a gap is open means confirming that no sibling RFC closes it (\S\ref{sec:dns}). 3GPP is the harder case: one user-visible behavior spans NGAP, F1AP, E1AP, RRC, and NAS at once, and error handling is routinely delegated across documents. 30 of the 530 findings mined agentically already turn on a normative reference outside the document they came from (\S\ref{sec:eval:rq1}). The interesting claim is not that a cross-document view finds more underspecifications. It admits a kind we cannot currently express, where no single document is silent yet the hand-off between two of them is unclosed: 
each is complete on its own terms but their composition is not. 

\paragraphb{State machine models:} We chose communicating state machines because they are the standard, minimal, and explainable account of a reactive protocol, and because the four categories fall directly out of them (\S\ref{sec:problem:categories}). The choice has a price, elaborated in Appendix~\ref{app:model}: data is folded into a finite symbolic state, and duration is not represented at all. Other lenses pay differently. An extended state machine adds context variables, guards, and update actions, so it models data flow alongside control~\cite{cheng1993efsm}, and its communicating variant is the direct successor to ours~\cite{bourhfir1998cefsm}. Under that lens several Effect findings stop being one merged state and become guard and update questions the model can check, such as which of two flow descriptors sharing an identifier wins. Timed automata~\cite{alur1994timed} can express what runs cannot: deadlines, timers, and retry policy, where a visible share of our divergences live. The tradeoff is significant: a richer lens yields a finer taxonomy but a harder extraction problem, the model now having to recover variables, guards, and clocks from prose.

\section{Conclusion}
\label{sec:conclusion}

Semantic underspecification is a structural blind spot in standards that pair formal syntax with natural-language prose. We addressed it by framing underspecification as a gap or fork in a partially specified communicating state machine, and built \sys to put that view in front of a language model as a scaffold. Across 36 5G procedures, our findings hold up under two independent checks, expert audit and differential testing. The same pipeline ports to DNS. 

The scaffold earns its place for a reason that outlasts any one model generation: more capable models can provide increased understanding, but not the criterion that decides what counts as an underspecification. Future models may simplify scaffold construction but will still need our state machine based (or related) classification. Further, our scaffolding is already built; our results indicate more powerful models do not unearth more underspecifications.

Finally, state machine models are classically used as an artifact to build implementations. Our paper turns this around, treating a formal model instead as a lens: this makes prose specifications analyzable at the scale they are written in complex modern protocols like 5G. Our taxonomy also fixes where formal verification takes over: \sys outputs each finding already localized, with competing completions and a discriminating test that model-based testing~\cite{eywa} can consume; one method finds where a model is worth building and the other verifies what happens once it is.

\newpage
\bibliographystyle{plain}
\bibliography{bibs}
\clearpage
\appendix
\section{A Communicating State-Machine Model}
\label{app:model}

This appendix expands the state-machine view of \S\ref{sec:problem:cfsm}. It gives the run semantics the main text abbreviates, restates gap and fork over that semantics, and states the test that separates underspecification from questions the model cannot answer.

\subsection{Communicating state machines}

A protocol of $n$ roles is a tuple $\mathcal{P} = \langle (S_i), (o_i), (M_{ij}), (\delta_i)\rangle$. For each role $i$, $S_i$ is a finite set of local states with initial state $o_i \in S_i$, and for each ordered pair $i \neq j$, $M_{ij}$ is the set of messages $i$ may send to $j$. A role acts by sending or receiving. Writing $-m$ for sending $m$ and $+m$ for receiving it, the local transition relation is
\[
  \delta_i \subseteq S_i \times A_i \times S_i,
  \qquad
  A_i = \{-m : m \in M_{ij}\} \cup \{+m : m \in M_{ji}\}.
\]

A global configuration $\gamma = (s, C)$ records every role's local state $s = (s_1,\dots,s_n)$ and the contents $C = (C_{ij})$ of the FIFO channels, each $C_{ij} \in M_{ij}^{*}$. The initial configuration $\gamma_0 = (o, \varepsilon)$ places every role in its initial state with empty channels. The step relation $\Rightarrow$ is the usual one: a send appends $m$ to $C_{ij}$ and advances the sender, and a receive consumes $m$ from the head of $C_{ji}$ and advances the receiver. A \emph{run} is a sequence $\gamma_0 \Rightarrow \gamma_1 \Rightarrow \cdots$, and the behavior of the protocol is the set $R$ of runs from $\gamma_0$.

\subsection{Specifications as partial machines}

A natural-language specification rarely determines a unique complete protocol. It may omit a case, permit several readings, or leave a progress obligation implicit, so we read it as inducing a partial, possibly nondeterministic relation $\delta_i^{\mathrm{spec}}$, whose runs form the set $R^{\mathrm{spec}}$. A conformant implementation is a deterministic refinement: it chooses concrete transitions within what the specification allows, $\delta_i^{\mathrm{impl}} \subseteq \delta_i^{\mathrm{spec}}$, under the chosen abstraction. Underspecification is any place where $R^{\mathrm{spec}}$ is left open, so that two conformant refinements may disagree.

\subsection{Local and temporal under-constraint}

Fix a reachable local state $s_i$ and an admissible action $a \in A_i$, and let
\[
  \Delta_i^{\mathrm{spec}}(s_i, a) = \{\, s_i' : (s_i, a, s_i') \in \delta_i^{\mathrm{spec}} \,\}.
\]
The step is well specified when $\lvert \Delta_i^{\mathrm{spec}}(s_i,a)\rvert = 1$, a \emph{gap} when the set is empty, and a \emph{fork} when it holds two or more elements. This is the local layer, and it splits a transition into its \textbf{Trigger}, whether a rule applies to $(s_i,a)$ at all, and its \textbf{Effect}, which next state and output the rule yields.

The step relation generates finite behaviors, but a protocol behavior is a set of complete runs. Under linear-time trace semantics a run property decomposes into a safety part, which finite prefixes are allowed, and a liveness part, what must eventually happen~\cite{alpern1985defining}. This is the temporal layer. \textbf{Ordering} is the safety side, constraining which interleavings of firings are allowed, and \textbf{Eventuality} is the liveness side, constraining obligations that no finite prefix can refute. Concurrency makes Ordering nontrivial, and loops make Eventuality nontrivial. The four components are exhaustive because a run-set is a temporal arrangement of firings over transitions with local content, and independent because each can vary while the others are fixed.

\subsection{Scope: specified, underspecified, or outside the model}

Not every behavioral question is one the model can answer. A question is \emph{in scope} exactly when its answer is a property of the run-set $R^{\mathrm{spec}}$: any two specifications that generate the same runs must answer it the same way. An in-scope question is \emph{specified} when one rule applies and \emph{underspecified} when it is a gap or a fork. It is \emph{out of model} when its answer can differ between two specifications that generate the very same runs, in which case no slot in the model records it. The one-line test is whether the answer could change without adding, removing, or reordering any run.

Three familiar questions fall outside it, one per standing assumption. Whether a choice is committed early or late is branching-time, and two specifications can resolve it differently while permitting the same runs. Whether a reply meets a deadline is real-time, and runs record order rather than duration. Distinctions the abstraction has already merged are invisible to it, so scope is always relative to that abstraction.

Two refinements decide the live cases. A gap counts only under \textbf{reachability}: some run must reach the configuration that exposes it, or the missing rule changes no behavior. And the model cannot judge \textbf{intent}: text loose enough to admit incompatible readings may be so by design rather than by oversight, so the model locates the opening and names its form while a human decides whether to close it or to document it.

\section{Key prompts (abridged)}
\label{app:prompts}

Each pipeline stage (\S\ref{sec:design}) is a single text-in, typed-JSON-out call. Figures~\ref{fig:prompt-mining}--\ref{fig:prompt-test} give the load-bearing parts of four of them; boilerplate, output schemas, and quoting limits are elided, and the full templates will be included in the planned code release. The CFSM framing is carried verbatim into the prompts, and the impact verdict is not asked of the model at all: it is computed by the deterministic engine over the fragment the model extracts. Figure~\ref{fig:prompt-naive} gives the naïve baseline of \S\ref{sec:eval:rq1}, which shares these inputs and output schema but withholds the framing.

\begin{figure}[t]
\begin{lstlisting}[style=prompt]
SYSTEM: You are a rigorous 3GPP/O-RAN control-plane analyst. You
model a specification as a communicating finite-state machine
(CFSM) and hunt for places where the spec leaves the run-set
under-constrained.

Model the procedure as communicating state machines exchanging
messages; a transition is (state, +message) -> (state', emitted
output), and an underspecification is an under-constraint of the
run-set they generate.
- Message/IE tables give syntax only. Assume syntax is valid.
- Behavioral text is where behavior should be pinned down,
  often silently.
- The 3GPP criticality mechanism (reject/ignore/notify) is
  syntax-level error handling, NOT a semantic validator; it does
  not resolve a behavioral gap.

[LOCAL pass] Mine ONLY the content of a single transition:
- Trigger: a reachable state where it is unclear whether a
  message is handled, more than one rule could apply, or a guard
  is never stated.
- Effect: a message is accepted but the resulting state
  update, response, or outcome is left open or reads two
  contradictory ways.

[TEMPORAL pass] Mine ONLY the arrangement of firings, and gate
first:
- Ordering applies only if the procedure admits concurrency /
  overlap.
- Eventuality applies only if a run can fail to make progress
  (loop, retry, pending obligation nothing forces to occur).
If neither precondition holds, return an empty findings array.

Reachability: report a finding only if the exposing
configuration is plausibly reachable; give a concrete run.
Prefer consequential over latent cases. Cite involved {messages,
ies, sections}; argue reachability inline.
\end{lstlisting}
  \caption{Stage~2, taxonomy-based mining (abridged): two scoped calls, local and temporal, the temporal one gated on applicability before mining.}
  \label{fig:prompt-mining}
\end{figure}

\begin{figure}[t]
\begin{lstlisting}[style=prompt]
SYSTEM: You decide the *form* of an underspecification: whether
the spec is silent (Gap) or permits several alternatives (Fork),
in CFSM run-set terms.

Consider the set Delta of behaviors the spec permits at the
under-constrained transition or temporal property:
- Gap (|Delta| = 0): no rule applies / no eventual obligation
  stated. A design defect; the specification itself should be
  fixed.
- Fork (|Delta| >= 2): more than one compliant behavior is
  allowed, possibly deliberate vendor latitude. The remedy is to
  test / document / tighten.
Intent: you can locate a fork but cannot decide whether it was
meant. Record the read in intent_note without treating it as
decisive.
\end{lstlisting}
  \caption{Stage~3, gap/fork triage (abridged): $|\Delta|=0$ (Gap) versus $|\Delta|\ge 2$ (Fork).}
  \label{fig:prompt-triage}
\end{figure}

\begin{figure}[t]
\begin{lstlisting}[style=prompt]
SYSTEM: You extract a small, faithful partial CFSM fragment from
the actual spec text and formalize each candidate behavior as a
transition. You never invent rules the spec does not support;
when something cannot be grounded in the text, leave that field
empty rather than guessing.

- A transition is (state, action) -> (next_state, outputs);
  action is +MSG/-MSG.
- specified_transitions: only transitions the spec actually
  pins down. For every pair of completions ending in different
  next_states, include >= 1 downstream transition that later
  reads those states and behaves differently for each -- the
  engine uses these to test whether the divergence is
  observable. A missing one makes a real divergence look latent.
- exposing_config.reach_trace: a concrete event sequence to
  the exposing configuration (the reachability witness). slot:
  the (process, state, action) the spec leaves open, with its
  form.
- Do NOT rank severity; it is computed mechanically
  downstream.
Field contract: decision in {accept, reject, ignore,
no_progress}; progresses is a boolean; outputs and reach_trace
are arrays of strings.
\end{lstlisting}
  \caption{Stage~4, CFSM model extraction (abridged): the model builds the fragment and formalizes the narratives; it never grades severity.}
  \label{fig:prompt-model}
\end{figure}

\begin{figure}[t]
\begin{lstlisting}[style=prompt]
SYSTEM: You turn a formal differential-test plan into a
concrete, executable interoperability test, preserving the
plan's branch expectations and discrimination method.

You are given a DETERMINISTIC plan (setup, trigger, branches,
discriminator) derived from the finding's CFSM model. Concretize
it: 1. Realize SETUP as concrete messages/IEs that reach the
exposing configuration. 2. Send the TRIGGER, with mandatory IEs
and the test-critical IEs that exercise the under-constraint. 3.
For EACH branch, give the concrete expected observable,
consistent with that branch's decision / expected_outputs /
next_state. Do not invent branches. 4. Implement the
DISCRIMINATOR exactly as planned: observe_decision,
downstream_probe, compare_notifications, or observe_progress.
Symbolic labels come from the model; map them to real message/IE
names. You must NOT alter the branch expectations or the
discrimination method.
\end{lstlisting}
  \caption{Stage~5, test concretization (abridged): the plan is fixed upstream; the model fills in concrete IE values only.}
  \label{fig:prompt-test}
\end{figure}

\begin{figure}[t]
\begin{lstlisting}[style=prompt]
SYSTEM: You are a rigorous 3GPP/O-RAN specification analyst.

You audit 3GPP/O-RAN RAN control-plane specifications for
underspecification. A specification is underspecified at a point
where its syntax is well defined (the messages and information
elements are declared, and it is clear what parses) but the
intended behavior is left open: the text does not pin down what
an implementation must actually do. Two implementations can then
both follow the specification and still behave differently.

# Task Read the specification context below and report every
underspecification you can find in it.

# Breadth & uniqueness Report all distinct underspecifications
you can justify (0..N; there is no target count). Ensure unique
titles; anchor each with the concrete section numbers it occurs
in and short supporting quotes.
\end{lstlisting}
  \caption{The naïve baseline for Stage~2 (\S\ref{sec:eval:rq1}), abridged. Its inputs and output schema match the miner of Figure~\ref{fig:prompt-mining}, but everything the framework contributes is withheld: no CFSM framing, no Trigger/Effect/Ordering/Eventuality taxonomy, no gap/fork form, and no reachability argument. Only a one-paragraph definition and an open-ended instruction to list every underspecification remain.}
  \label{fig:prompt-naive}
\end{figure}

\section{End-to-end testbed validation}
\label{app:e2e}

The component tests of \S\ref{sec:impl:harness} isolate the control-plane logic under test. To confirm that the divergences also surface in a running network, we reproduced six of them end to end. Two RAN implementations, OpenAirInterface and srsRAN, ran on a workstation with an Intel Xeon W5-3435X and 64\,GB of DDR5 memory under Ubuntu 22.04, each attached to its 5G core. For each finding we encoded the trigger with \textsf{pycrate}, filled abstract fields such as transaction identifiers from the live session, injected it over SCTP with \textsf{pysctp}, and read the response from console logs and returned messages. Table~\ref{tab:e2e} lists the six reproduced cases; each corresponds to a confirmed divergence in Appendix~\ref{app:divergences}.

\begin{table}[h]
\setlength{\tabcolsep}{4pt}
\begin{tabular}{@{}p{0.66\columnwidth}p{0.22\columnwidth}@{}}
\toprule
\textbf{Underspecification exercised} & \textbf{Protocol} \\
\midrule
Inconsistent CU/DU UE-ID pair in a partial F1 RESET       & F1AP \\
Duplicate or already-active PDU Session ID during concurrent setup & NGAP \\
PDU Session ID reused while its release is still pending    & NGAP \\
Unknown PDU Session ID in a RELEASE COMMAND                 & NGAP \\
Downlink NAS identifiers resolving no live UE association   & NGAP \\
E2 Setup outcome with no matching live transaction          & E2AP \\
\bottomrule
\end{tabular}
\caption{The six divergences reproduced on the live testbed.}
\label{tab:e2e}
\end{table}

\section{Evaluated procedures}
\label{app:procedures}

The 36 evaluated procedures, those that both OpenAirInterface and srsRAN implement, are listed below by specification.

\paragraphb{GTP-U (TS~29.281).} Echo Request, Echo Response, G-PDU.

\paragraphb{NGAP (TS~38.413).} PDU Session Resource Setup, Modify, and Release; Initial Context Setup; UE Context Release; Paging; Initial UE Message; Downlink and Uplink NAS Transport; NG Setup; UE Radio Capability Info Indication.

\paragraphb{E1AP (TS~38.463).} gNB-CU-UP E1 Setup; Bearer Context Setup, Modification, and Release.

\paragraphb{F1AP (TS~38.473).} Reset; F1 Setup; gNB-DU and gNB-CU Configuration Update; UE Context Setup, Release, and Modification; Initial UL, DL, and UL RRC Message Transfer; Paging.

\paragraphb{E2AP.} RIC Subscription, RIC Subscription Delete, RIC Indication, RIC Control, E2 Setup.

\paragraphb{E2SM-KPM.} Periodic Report; common condition-based UE-level measurement.

\section{Effectiveness: instruments and results}
\label{app:rq1}

\paragraphb{Reviewability checklist.} A blinded finding is scored on six concrete criteria: (1) it is bounded to the cited material; (2) it names both a message and an information element; (3) it cites two or more sections; (4) it supplies two or more evidence quotes; (5) it has at least six concrete anchors in total; and (6) it is judged highly actionable. A finding meeting at least four criteria is \emph{reviewer-ready}. The first five criteria are deterministic; only actionability uses a judge. On a deterministic-only variant of the score the pipeline is reviewer-ready for 81\% of findings against 59\% for naïve.

\begin{table}[t]
  \centering
  \setlength{\tabcolsep}{4pt}
  \begin{tabular}{@{}lcc@{}}
    \toprule
    \textbf{gpt 5.4 nano / mini / full} & \textbf{\sys} & \textbf{Naïve} \\
    \midrule
    Temporal reach (judge 2)   & 50 / 45 / 51 & 44 / 29 / 33 \\
    Temporal reach (judge 1)   & 60 / 45 / 53 & 37 / 25 / 30 \\
    Messages per 100 findings  & 75 / 100 / 53 & 63 / 68 / 34 \\
    Repeats per finding         & .20 / .15 / .27 & .33 / .18 / .44 \\
    Cite two or more sections  & 95 / 95 / 100 & 33 / 100 / 100 \\
    \bottomrule
  \end{tabular}
  \caption{Capability progression on a five-procedure microbenchmark (\S\ref{sec:eval:rq1}). Rows are percentages, except repeats per finding (a $[0,1]$ rate).}
  \label{tab:rq1-prog}
\end{table}

\paragraphb{Agentic mining, per protocol and per model.} Of the 530 findings behind Table~\ref{tab:noharness}, gpt-5.6-sol produced 269 and claude-opus-5 261, judged genuine at 41\% and 51\%. The spread by protocol is far wider: E2AP best at 62 of 97, F1AP worst at 20 of 86, where 57 of 86 are deliberate policy freedom or plain optionality. Delegation is the other recurring misread: one finding calls GTP-U path failure undefined while quoting the clause that hands it to a procedure in another document.

\paragraphb{What agentic mining misses.} Table~\ref{tab:noharness} counts what agentic mining reports; a second audit asks what it fails to report. We took four procedures whose pipeline reports align cleanly with the chat outputs, F1AP UE Context Setup, NGAP PDU Session Resource Setup, E2AP RIC Subscription, and E2SM-KPM Periodic Report, and cross-checked each of the 36 pipeline findings against both models' output, counting it fully recovered only when the chat finding could substitute for it in a review package. gpt-5.6-sol recovers 12 fully and 17 partially and misses 7; claude-opus-5 recovers 14 and 13 and misses 9. Taking the better of the two per finding, 14 of 36 are fully recovered, 22 are at best partial, and 5 of those are missed by both. The misses give one example per category of \S\ref{sec:problem:categories}: an SRB ID of 0 the receive procedure never gives meaning to and a repeated QoS Flow Identifier in one request; a Serving NID the receive path need not store and ordinary SRB setup success that is never reported; a conditional-handover replacement whose prepared target is absent; and a periodic-report trigger nothing forces to fire.

\paragraphb{Redundancy metric.} Clustering findings by the protocol locus they name, a repeat is any finding that only re-expresses a missing-rule pattern already recorded at that locus, the ``optional IE absent, so behavior is undefined'' template instantiated per field. The raw count is 155 for the pipeline against 269 for naïve; the \emph{repeats per finding} row of Table~\ref{tab:rq1-prog} divides it by the backend's number of findings, and lower means less templating. Both are deterministic and use no judge.

\paragraphb{Reach.} A finding is \emph{cross-step}, the \emph{temporal reach} row of Table~\ref{tab:rq1-prog} and the \emph{cross-step} row of Table~\ref{tab:rq1}, when a blinded judge labels it Ordering or Eventuality (interleaving, retry, timeout, liveness, late or duplicate handling) rather than a single local step.

\paragraphb{Iterative few-shot.} The few-shot sub-experiment is 18 runs: three backends, three procedures, treatment and control, four rounds each. Treatment seeds round $N{+}1$ with prior rounds' curated findings; control runs the identical loop with no exemplars. The similarity and diversity gaps of \S\ref{sec:eval:rq1} are measured on the raw pre-deduplication generations, and hold in 6 of 8 and 7 of 8 pairs respectively. Both treatment and control mode-collapse: about 80\% of what rounds 2 to 4 generate is already a duplicate. Count-based metrics are run-to-run noisy and we draw no conclusions from them.

\section{Expert audit and repair patterns}
\label{app:rq2}

\paragraphb{Repair patterns.} The 40 gaps among the 145 unanimous proposals cluster into three fixes: \emph{complete the postcondition} (24 Effect gaps, a \texttt{shall} rule fixing the next state, output, or cleanup), \emph{complete the receive table} (13 Trigger gaps, requiring accept, reject, ignore, or error for the omitted input and state), and \emph{close the lifecycle} (3 Eventuality gaps, a timer or retry bound and a terminal state). The 105 forks cluster into four: \emph{pin the effect or expose the choice} (71 Effect forks), \emph{make guards mutually exclusive} (15 Trigger forks), \emph{define atomicity and happens-before} (17 Ordering forks), and \emph{require termination} (2 Eventuality forks).

\paragraphb{Drafting templates.} Each pattern maps to a normative template:
\begin{denseitemize}
\item \emph{Gap / Trigger.} ``Upon receiving $M$ in state $S$, the node shall $A$; if $C$ is false, it shall $B$ and enter $S'$.''
\item \emph{Gap / Effect.} ``After accepting $M$, the node shall update $X$, emit $Y$, and remove or retain $Z$.''
\item \emph{Fork / Effect.} ``The node shall use behavior $A$. If $B$ is supported, support shall be negotiated by $C$; absence of $C$ means $A$.''
\item \emph{Fork / Ordering.} ``$A$ shall complete before $B$; simultaneous events resolve in favor of $A$; failure rolls back $X$ but retains $Y$.''
\item \emph{Eventuality.} ``If no response arrives within $T$, the node shall retry at most $N$ times and then enter terminal state $S_t$ while emitting $E$.''
\end{denseitemize}

\section{Confirmed divergences by protocol}
\label{app:divergences}

The 60 confirmed divergences of \S\ref{sec:eval:rq3} are listed below, grouped by protocol. Each entry names the underspecification, then how OpenAirInterface (OAI) and srsRAN resolve it with a code location, and the severity as \emph{observed\,/\,predicted}: CA (Conflicting Actions), IIS (Inconsistent Internal States), or IN (Inconsistent Notifications). Locations are file:line in the pinned revisions of \S\ref{sec:impl:harness}.

\paragraphb{NGAP (gNB peer).}
\begin{denseitemize}
\item \textbf{Conflicting UE-ID pair in DL NAS Transport.} OAI keys by RAN-UE-ID only and retains the stale context (\url{ngap_gNB_nas_procedures.c:334}); srsRAN releases the old context and emits an Error Indication. \emph{CA\,/\,CA}
\item \textbf{Unknown PDU Session ID in RELEASE COMMAND.} OAI skips the ID, keeping the UE (\url{rrc_gNB_NGAP.c:1922}); srsRAN rejects the whole command and releases the entire UE (\url{pdu_session_resource_release_routine.cpp:194}). \emph{CA\,/\,CA}
\item \textbf{Duplicate / already-active PDU Session ID in SETUP.} OAI fails all, then admits the fresh ID (\url{rrc_gNB_NGAP.c:908}); srsRAN fails the duplicate and admits the fresh ID atomically (\url{ngap_validators.cpp:42}). \emph{CA\,/\,CA}
\item \textbf{One infeasible QoS flow among feasible ones in MODIFY.} OAI keeps the session, skipping the bad flow (\url{rrc_gNB_NGAP.c:1082}); srsRAN fails the whole PDU session (\url{up_resource_manager_helpers.cpp:388}). \emph{CA\,/\,CA}
\item \textbf{MODIFY releasing the last flow of a session.} OAI reclaims the emptied DRB, reporting success (\url{rrc_gNB_NGAP.c:1029}); srsRAN rejects the whole request (\url{up_resource_manager_helpers.cpp:174}). \emph{CA\,/\,IIS}
\item \textbf{Preferred UP security with no mandated treatment.} OAI emits no Security Result; srsRAN reports protection not-performed. \emph{IN\,/\,CA}
\item \textbf{Two QoS flows with the same QFI.} OAI keeps the first descriptor, srsRAN the last. \emph{IIS\,/\,CA}
\item \textbf{NG SETUP FAILURE Time-to-Wait, no retry policy.} OAI reads only the cause and never retries (\url{ngap_gNB_handlers.c:96}); srsRAN waits the delay and retries (\url{ng_setup_procedure.cpp:120}). \emph{IIS\,/\,IIS}
\item \textbf{Late / duplicate NG SETUP RESPONSE.} OAI re-counts, incrementing the association counter (\url{ngap_gNB_handlers.c:305}); srsRAN drops it via the transaction sink (\url{ng_setup_procedure.cpp:60}). \emph{IIS\,/\,CA}
\item \textbf{RELEASE COMMAND for an unknown UE context.} OAI silently drops (\url{ngap_gNB_handlers.c:878}); srsRAN drops and emits an Error Indication (\url{ngap_impl.cpp:702}). \emph{IN\,/\,CA}
\item \textbf{Replayed RELEASE COMMAND after release scheduled.} OAI silently drops; srsRAN emits Error Indication \texttt{interaction-with-other-proc} (\url{ngap_impl.cpp:722}). \emph{IN\,/\,IN}
\item \textbf{ICS FAILURE cause on a DU-originated setup failure.} OAI's CU handler is unimplemented and aborts (\url{f1ap_cu_ue_context_management.c:65}); srsRAN emits ICS Failure with \texttt{protocol/unspecified} (\url{initial_context_setup_routine.cpp:62}). \emph{CA\,/\,IN}
\item \textbf{UE Context RELEASE REQUEST cause selection.} OAI hard-codes \texttt{radio-connection-lost} (\url{rrc_gNB.c:2681}); srsRAN maps the DU's F1AP cause (\url{f1ap_cause_converters.cpp:29}). \emph{IN\,/\,IN}
\item \textbf{RELEASE COMPLETE payload composition.} OAI never attaches User Location Information, srsRAN always (\url{ue_context_release_routine.cpp:55}). \emph{IN\,/\,IN}
\item \textbf{Invalid UE security capabilities plus PDU-session setup.} OAI continues into per-session setup; srsRAN emits ICS Failure naming the session. \emph{CA\,/\,CA}
\item \textbf{Unknown local UE ID on Initial Context Setup.} OAI emits ICS Failure; srsRAN emits Error Indication \texttt{unknown-local-UE-ID}. \emph{IN\,/\,CA}
\item \textbf{RELEASE COMMAND NAS forwarding vs teardown ordering.} OAI defers cleanup to the release response; srsRAN releases the NAS first. \emph{IIS\,/\,IIS}
\item \textbf{Unsupported REROUTE NAS REQUEST.} OAI's decoder default aborts the CU (\url{ngap_gNB_decoder.c}); srsRAN drops the message gracefully. \emph{CA\,/\,IIS}
\item \textbf{Uplink NAS before the AMF association exists.} OAI forwards without the AMF-ID guard; srsRAN drops until the association exists. \emph{CA\,/\,CA}
\item \textbf{Duplicate Initial UE Message while first association pending.} OAI can emit a second Initial UE Message; srsRAN suppresses it. \emph{CA\,/\,CA}
\item \textbf{Released-list duplicate canonicalization.} OAI deduplicates by internal session; srsRAN drops duplicate entries. \emph{IN\,/\,IN}
\end{denseitemize}

\paragraphb{F1AP (CU\,$\leftrightarrow$\,DU).}
\begin{denseitemize}
\item \textbf{GNB-CU CONFIGURATION UPDATE with PCI omitted.} OAI flattens PCI to zero and aborts on an assertion (\url{f1ap_interface_management.c:1751}); srsRAN projects the CGI only and ACKs (\url{cu_configuration_procedure.cpp:53}). \emph{CA\,/\,IIS}
\item \textbf{SRB ID = 0 in UE CONTEXT MODIFICATION.} OAI passes SRB0 to the RLC guard and asserts (\url{mac_rrc_dl_handler.c:209}); srsRAN maps 0 to SRB0 and returns success. \emph{CA\,/\,CA}
\item \textbf{F1 SETUP REQUEST with the optional gNB-DU Name absent.} OAI proceeds to Setup Response (\url{f1ap_interface_management.c:790}); srsRAN fails setup (\url{f1_setup_procedure.cpp:37}). \emph{CA\,/\,CA}
\item \textbf{No timeout when F1 Setup is unanswered.} OAI polls forever (\url{gnb_config.c:2699}); srsRAN times out at 3000\,ms (\url{f1ap_du_setup_procedure.cpp:35}). \emph{CA\,/\,CA}
\item \textbf{DU state after F1 SETUP FAILURE with Time-to-Wait.} OAI calls \texttt{exit(1)} (\url{mac_rrc_dl_handler.c:196}); srsRAN waits and retries (\url{f1ap_du_setup_procedure.cpp:128}). \emph{CA\,/\,IIS}
\item \textbf{Unsupported new gNB-CU UE F1AP ID in DL RRC transfer.} OAI keeps the old binding (\url{f1ap_rrc_message_transfer.c:373}); srsRAN rebinds to the new ID (\url{f1ap_du_impl.cpp:270}). \emph{IIS\,/\,IIS}
\item \textbf{Paging DRX / Priority optional hints.} OAI drops both for the cell default (\url{f1ap_paging.c:214}); srsRAN preserves DRX for paging-frame selection (\url{paging_scheduler.cpp:144}). \emph{IIS\,/\,IIS}
\item \textbf{Full-vs-delta CellGroupConfig when a source config exists.} OAI omits Full Configuration (\url{mac_rrc_dl_handler.c:615}); srsRAN emits it (\url{ue_configuration_procedure.cpp:351}). \emph{IN\,/\,IN}
\item \textbf{Measurement-gap creation discretionary after MeasConfig.} OAI creates no gap (\url{mac_rrc_dl_handler.c:683}); srsRAN derives a gap from SMTC1 (\url{du_meas_config_manager.cpp:99}). \emph{IN\,/\,IN}
\item \textbf{Transmission stop/restart transition point undefined.} OAI defers via a timeout, staying schedulable (\url{mac_rrc_dl_handler.c:892}); srsRAN disconnects DRBs before the Response (\url{du_ue_controller_impl.cpp:274}). \emph{IIS\,/\,IIS}
\item \textbf{RRC delivery not ordered against teardown in RELEASE COMMAND.} OAI runs a delivery-blind 100\,ms timer (\url{mac_rrc_dl_handler.c:973}); srsRAN awaits the delivery callback or 120\,ms (\url{f1ap_du_ue_context_release_procedure.cpp:64}). \emph{IIS\,/\,IIS}
\item \textbf{Per-item semantics of a partial-F1 RESET.} OAI aborts on any partial RESET (\url{f1ap_du_interface_management.c}); srsRAN resolves the UE and ACKs (\url{f1ap_du_impl.cpp:156}). \emph{CA\,/\,IIS}
\item \textbf{Unmodeled optional IEs in GNB-CU CONFIGURATION UPDATE.} OAI's decoder models a subset and drops the message; srsRAN ignores the IE and ACKs (\url{f1ap_cu_impl.cpp}). \emph{IIS\,/\,IIS}
\item \textbf{Dedicated-SI list in GNB-DU CONFIGURATION UPDATE.} OAI default-rejects and drops (\url{f1ap_interface_management.c:1584}); srsRAN ignores contents and ACKs (\url{f1ap_cu_impl.cpp:266}). \emph{IIS\,/\,IIS}
\item \textbf{Receive-state precondition for GNB-DU CONFIGURATION UPDATE.} OAI requires a bound DU, dropping otherwise (\url{rrc_gNB_du.c:987}); srsRAN ACKs unconditionally. \emph{IIS\,/\,CA}
\item \textbf{CU System Information in F1 SETUP RESPONSE.} OAI applies the CU-owned SIBs at the DU (\url{f1ap_interface_management.c:662}); srsRAN reads only the CGI. \emph{IIS\,/\,IIS}
\item \textbf{Old gNB-DU UE F1AP ID on release, RRC container absent.} OAI retains the old context; srsRAN schedules its removal (\url{f1ap_du_impl.cpp:187}). \emph{IIS\,/\,IIS}
\item \textbf{Paging Cell List not validated against served cells.} OAI never inspects the list; srsRAN validates each cell and pages none if unmatched (\url{f1ap_du_impl.cpp:587}). \emph{IIS\,/\,IIS}
\item \textbf{Absent DU-to-CU RRC container on reestablishment.} OAI has no reject gate and proceeds (\url{rrc_gNB.c:1435}); srsRAN rejects the reestablishment. \emph{CA\,/\,CA}
\end{denseitemize}

\paragraphb{E2AP (E2 node; OAI's E2 agent is FlexRIC).}
\begin{denseitemize}
\item \textbf{RIC Control Ack Request extension ordinal.} FlexRIC asserts the value is \texttt{ACK} and aborts (\url{msg_handler_agent.c:246}); srsRAN's decoder rejects the extension value (\url{e2ap.h:2783}). \emph{CA\,/\,CA}
\item \textbf{Whether the RAN Function ID must match on subscription delete.} FlexRIC ignores the key miss and keeps the subscription (\url{msg_handler_agent.c:200}); srsRAN keys by requestor only and erases it (\url{e2_subscription_manager_impl.cpp:88}). \emph{IIS\,/\,CA}
\item \textbf{Optional Call Process ID in RIC CONTROL ACKNOWLEDGE.} FlexRIC sets it to \texttt{NULL} (\url{msg_handler_agent.c:270}); srsRAN echoes the request field (\url{e2_ric_control_procedure.cpp:72}). \emph{IN\,/\,CA}
\item \textbf{No retry/abort bound when E2 Setup is unanswered.} FlexRIC retransmits forever on a 3\,s timer (\url{e2_agent.c:474}); srsRAN's transaction expires (\url{e2_setup_procedure.cpp:80}). \emph{CA\,/\,IN}
\item \textbf{Subscription DELETE for no matching subscription.} FlexRIC returns success; srsRAN returns DELETE FAILURE. \emph{CA\,/\,CA}
\item \textbf{Unsupported RIC SUBSCRIPTION MODIFICATION REQUEST.} FlexRIC's dispatch guard aborts the agent; srsRAN declines without crashing. \emph{CA\,/\,CA}
\item \textbf{Post-setup state for functions omitted from the Accepted list.} FlexRIC still services them via its plugin registry; srsRAN promotes only Accepted-list functions. \emph{CA\,/\,CA}
\end{denseitemize}

\paragraphb{E1AP (CU-CP\,$\leftrightarrow$\,CU-UP).}
\begin{denseitemize}
\item \textbf{No progress rule if CU-CP never answers CU-UP E1 Setup.} OAI has no response timer and pends (\url{e1ap.c:500}); srsRAN sets a 3000\,ms deadline, then a fatal setup error (\url{e1ap_cu_up_setup_procedure.cpp:33}). \emph{CA\,/\,CA}
\item \textbf{CU-CP cleanup after BEARER CONTEXT RELEASE COMPLETE.} OAI does no cleanup on Complete, reclaiming later (\url{rrc_gNB.c:3356}); srsRAN removes the E1 UE context immediately (\url{bearer_context_release_procedure.cpp:43}). \emph{IIS\,/\,IIS}
\item \textbf{Admitted PDU session with zero successful child DRBs.} OAI reports both DRBs successful; srsRAN reports parent success with DRB failures. \emph{CA\,/\,CA}
\item \textbf{Duplicate CU-CP UE E1AP ID in Bearer Context Setup.} OAI reuses the CU-UP UE ID; srsRAN allocates a fresh one, keeping two contexts. \emph{IIS\,/\,CA}
\item \textbf{Contradictory inactivity-monitoring configuration.} OAI returns SETUP RESPONSE; srsRAN returns SETUP FAILURE. \emph{CA\,/\,CA}
\item \textbf{Rejection granularity for an unsupported child resource.} OAI reports the child successful; srsRAN reports a parent failure. \emph{CA\,/\,IIS}
\item \textbf{Storage scope when multiple PLMNs carry distinct child lists.} OAI has no NR-CGI support field and loses that scope; srsRAN retains PLMN, slice, and NR-CGI per item. \emph{IIS\,/\,CA}
\item \textbf{Success response when preferred UP confidentiality is not performed.} OAI omits the Security Result; srsRAN emits \texttt{not-performed}. \emph{IN\,/\,CA}
\item \textbf{Local reaction to a wrong-role first CU-UP Setup PDU.} OAI's global dispatch forwards it to the CU-CP path; srsRAN's role-scoped ingress drops it. \emph{CA\,/\,IN}
\end{denseitemize}

\paragraphb{GTP-U.}
\begin{denseitemize}
\item \textbf{Header-only container G-PDU with a zero-length T-PDU.} OAI guards callbacks on size and delivers nothing (\url{gtp_itf.cpp:1188}); srsRAN extracts the empty T-PDU and calls up (\url{gtpu_tunnel_nru_rx_impl.h:52}). \emph{IIS\,/\,IIS}
\item \textbf{Unknown comprehension-required extension header.} OAI skips the extension, delivering the T-PDU (\url{gtp_itf.cpp:1217}); srsRAN's comprehension check drops the G-PDU. \emph{CA\,/\,IN}
\end{denseitemize}

\paragraphb{E2SM-KPM.}
\begin{denseitemize}
\item \textbf{Boolean guard over a Style-4 Matching Condition list.} OAI evaluates each condition, suppressing the report on an empty one (\url{ran_func_kpm.c:338}); srsRAN's matcher is a stub, reporting every tracked UE (\url{e2sm_kpm_du_meas_provider_impl.cpp:223}). \emph{CA\,/\,IIS}
\item \textbf{Receive rule for an unsupported Measurement ID.} OAI admits, then aborts during reporting; srsRAN rejects at admission. \emph{CA\,/\,CA}
\end{denseitemize}

\section{Additional 5G case studies}
\label{app:cases}

\paragraphb{A single message that aborts a base station.} The physical cell identity is optional in GNB-CU CONFIGURATION UPDATE. Omitted for an already-configured cell, OAI flattens it to zero, compares it with the configured value, and reaches an assertion that ends the process; srsRAN keeps its local value and acknowledges. One legal message thus aborts the OAI distributed unit, a remotely triggerable denial of service.

\paragraphb{An under-constrained measurement filter.} E2SM-KPM Report Style~4 scopes UE-level measurements by a \textsf{Matching Condition} list, but fixes no Boolean guard over it: neither how conditions combine nor what a condition matching no device means. OAI evaluates item by item and suppresses the whole report once any condition matches nothing; srsRAN's matcher does not evaluate the list and returns every device it tracks. The empty-match and combination semantics are the gap, and srsRAN's non-evaluation compounds it: a subscription meant to scope one slice can receive measurements for every device tracked.

\paragraphb{A final message that one implementation may drop.} On UE Context Release the specification does not order RRC delivery against teardown. OAI arms a delivery-blind 100\,ms timer and can delete the UE before the final RRC message is delivered; srsRAN waits for the delivery callback or 120\,ms. A last RRCRelease carrying a redirect can therefore be lost on OAI under load: the network treats the UE as released while the UE never receives its redirect.

\paragraphb{Silent QoS-policy divergence.} A PDU session setup with two flows carrying the same QoS identifier has no receive rule. OAI commits the first descriptor and srsRAN the last, and both return success, so the same request installs different QoS on the two implementations. A UE's flow is enforced at one QoS class on OAI and another on srsRAN, and neither the core network nor the operator is told the two disagree.

\paragraphb{A subscription one implementation keeps and the other erases.} On a RIC subscription delete carrying a mismatched function identifier, FlexRIC does not find the entry yet returns success, leaving the subscription live; srsRAN erases it. Both first responses are identical, so a static reading would conclude the implementations agree. The desync appears only later, when a controller that believes a subscription gone still receives indications from FlexRIC.

\section{DNS Generalization}
\label{app:dns}

This appendix backs the DNS case study of \S\ref{sec:dns}: the adaptation footprint, the RFC bases and per-server outcomes behind the findings we present, and the family-level filter we apply before treating a divergence as a finding. We report these findings as case studies and do not claim a complete count.

\paragraphb{Adaptation footprint.} The tracked \sys workflow is about 4{,}770 lines of code and 777 lines of prompt text. Adapting it to DNS replaced about 26 lines of code, roughly 0.5\%, and 40 to 50 lines of prompt text, roughly 5 to 6\%, for some 1.2 to 1.4\% of the tracked workflow overall. The workflow edits are mechanical: thread a domain flag through the existing stages, switch prompt rendering by domain, add DNS metadata fields and entry points, and drop the intent-judgment language from triage. The larger effort was additive: cross-RFC reference-graph ingestion, corpus indexing, and DNS planning, validation, and audit are new modules, not edits to the 5G path. We run the findings differentially on the Ferret testbed~\cite{kakarla2022scale} across BIND~\cite{bind9}, Knot~\cite{knotdns}, Technitium~\cite{technitium}, YADIFA~\cite{yadifa}, and Hickory DNS~\cite{hickorydns}.

\paragraphb{QCLASS NONE in a standard query (Trigger, Gap).} RFC~6895 registers class value 254 as \textsf{QCLASS NONE}, a value meaningful in queries and updates~\cite{rfc6895}, and RFC~2136 gives it a meaning only inside UPDATE prerequisite and update sections~\cite{rfc2136}. No RFC defines how an authoritative server processes an \textsf{OPCODE=QUERY} question carrying it, and RFC~9499 still describes a query as a name, type, and class without a rule for this class~\cite{rfc9499}. The message decodes, so the opening is at transition selection. On a query for the name \textsf{www.campus.edu.} of type A, BIND returns \textsf{FORMERR}, Knot \textsf{REFUSED}, Technitium \textsf{REFUSED} with recursion-available set, YADIFA \textsf{NOTIMP}, and Hickory DNS \textsf{NOERROR} with the \textsf{IN} record \textsf{192.0.2.50}, four incompatible outcomes for one syntactically valid message. Ordinary clients query class \textsf{IN}, so the case bears on robustness rather than a demonstrated production failure. Hickory DNS returning \textsf{IN} data for a non-\textsf{IN} question is the sharper edge, since DNS classes are separate namespaces~\cite{rfc6895}, though we do not show that answer entering a cache.

\paragraphb{SOA serial increment granularity (Effect, Fork).} RFC~2136 \S3.6 requires an accepted update that does not itself change the SOA serial to advance it automatically before the change becomes visible, and it lists several permissible triggering events without mapping one update message or one mutation to one serial unit~\cite{rfc2136}. RFC~1982 \S3.1 defines advancement as addition of any positive integer~\cite{rfc1982}, and RFC~3007 \S1.1 requires an increment but no scheme~\cite{rfc3007}. Measuring each server's serial immediately before and after one accepted update that adds an A record, adds a TXT record, and deletes an A record: BIND and Knot advance from 3 to 4, a delta of one; Technitium advances from 6 to 9, a delta of three; YADIFA rejects the update and is not a witness. The before measurement removes a setup confound, since Technitium's import had already advanced its serial to 6. Ordinary resolution is unaffected, since a secondary checks only whether the serial is newer. Stateful automation is not: a client that reads the serial, updates, and reuses the predicted value in a later exact-serial prerequisite succeeds under a per-transaction policy and fails under a per-mutation one.

\paragraphb{Duplicate add with a different TTL (Effect, Fork).} Here the opening is a conflict between clauses rather than silence. For a zone record \textsf{host1.campus.edu.} of type A at TTL 300, an accepted update adds the same owner, class, type, and data at TTL 60. RFC~2136 \S1.1.1 excludes TTL from RR equality and \S2.5.1 says duplicate additions are silently ignored, which preserve TTL 300, while \S3.4.2.2 says a duplicate-RDATA add replaces the zone RR with the update RR, which installs TTL 60~\cite{rfc2136}. BIND and YADIFA install 60; Knot and Technitium keep 300. RFC~2181 \S5 forbids the mixed-TTL and duplicate state but does not select which TTL survives this update~\cite{rfc2181}. The normative text points two ways, so implementations diverge; we report it as an effect fork, with the caveat that its source is a contradiction, so no implementation can honor both clauses at once. This case carries the clearest operational consequence. Operators lower a TTL before changing an address, so if the duplicate add is ignored, a cutover takes effect later than the management system expects and a desired-state controller reissuing the change never converges on the preserving servers. Deleting and re-adding the RRset avoids this, but the portability gap remains.

\paragraphb{Filtering against the RFC family.} Because a DNS contract spans several RFCs, some candidate divergences close once the whole family is read; we exclude them. Whether a same-message add and delete of one record leaves it present or absent is fixed by the in-order processing rule of RFC~2136 \S3.4.2~\cite{rfc2136}, so a server that diverges there is nonconformant rather than exercising a gap. We also do not count gaps an RFC itself flags as undefined, such as the semantics of a wildcard-owned NS RRset, which RFC~4592 \S4.2 leaves explicitly open~\cite{rfc4592}; these are real but already acknowledged, so they are weak as discoveries.

\end{document}